\documentclass{aa}
\usepackage{txfonts}
\usepackage{graphicx}
\usepackage{setspace}
\usepackage{natbib}
\usepackage{sidecap}
\bibpunct{(}{)}{;}{a}{}{,}
\usepackage{color}
\usepackage{subfigure}
\usepackage{floatflt}
\usepackage{rotating}
\usepackage{multirow}
\usepackage{aalongtable, lscape, afterpage}
\usepackage{supertabular}

\newcommand{\rul}{\rule[-2.50mm]{0mm}{1mm}}
\newcommand{\comment}[1]{}

\begin{document}


\newcommand{\st}[1]{\mathrm{#1}} 
\newcommand{\bld}[1]{\textbf{#1}}
\newcommand{\pow}[2]{$\st{#1}^{#2}$}
\newcommand{\grad}{\hspace{-0.15em}\r{}}
\newcommand{\lm}{\lambda}

\newcommand{\average}[1]{\left\langle #1 \right\rangle}

\newcommand{\h}{~h_{71}~}
\newcommand{\hinv}[1]{~h_{71}^{#1}~}
\newcommand{\eq}[1]{(\ref{eq-#1})}
\newcommand{\wrt}{with respect to\ }
\newcommand{\mms}{\frac{M_{\odot}}{M}}
\newcommand{\mpy}{\st{M}_{\odot}~\st{yr}^{-1}}
\newcommand{\ct}{t_{\st{cool}}}
\newcommand{\rc}{r_{\st{cool}}}
\newcommand{\tvir}{T_{\st{vir}}}
\newcommand{\rvir}{R_{\st{500}}}
\newcommand{\rcore}{r_{\st{core}}}
\newcommand{\mvir}{M_{\st{500}}}
\newcommand{\mbh}{M_{\st{BH}}}
\newcommand{\ms}{\st M_{\odot}}
\newcommand{\ls}{L_{\odot}}
\newcommand{\lxb}{L_{\st {Xb}}}
\newcommand{\lx}{L_{\st {X}}}
\newcommand{\lxc}{L_{\st {X, core}}}
\newcommand{\flxc}{{f}_{\lxc}}
\newcommand{\lxrc}{L_{\st {X, rcool}}}
\newcommand{\flxrc}{{f}_{\lxrc}}
\newcommand{\lt}{L_{\st{X}}{\st -}\tvir}
\newcommand{\lr}{L_{\st {R}}}
\newcommand{\lbcg}{L_{\st{BCG}}}
\newcommand{\Dl}{D_{\st{l}}}
\newcommand{\hiflux}{\textit{HIFLUGCS}}
\newcommand{\chandra}{\textit{Chandra}}
\newcommand{\vla}{\textit{VLA}}
\newcommand{\gmrt}{\textit{GMRT}}
\newcommand{\atca}{\textit{ATCA}}
\newcommand{\XMM}{\textit{XMM-Newton}}
\newcommand{\einstein}{\textit{Einstein}}
\newcommand{\asca}{\textit{ASCA}}
\newcommand{\rosat}{\textit{ROSAT}}
\newcommand{\mdr}{\dot{M}_{\st{classical}}}
\newcommand{\smdr}{\dot{M}_{\st{spec}}}

\title{The $\lt$ relation in galaxy clusters: \\Effects of radiative
  cooling and AGN heating}

\author{Rupal~Mittal\inst{1,2} \and Amalia~Hicks\inst{3} \and
  Thomas~H.~Reiprich\inst{2} \and Vera Jaritz\inst{2}
  \institute{Center of Imaging Science, Rochester Institute of
    Technology, 54 Lomb Memorial Drive, Rochester, NY, USA 14623 \and
    Argelander-Institut f\"ur Astronomie, Auf dem H\"ugel 71, 53121
    Bonn, Germany \and Michigan State University, Physics and
    Astronomy Dept., East Lansing, MI 48824-2320, USA}}

\date{Received/Accepted}

\abstract{We present a detailed investigation of the X-ray
  luminosity~($\lx$)-gas temperature~($\tvir$) relation of the
  complete X-ray flux-limited sample of the 64 brightest galaxy
  clusters in the sky ($\hiflux$). We study the influence of two
  astrophysical processes, active galactic nuclei~(AGN) heating and
  intracluster medium~(ICM) cooling, on the $\lt$ relation,
  simultaneously for the first time. We employ homogeneously
  determined gas temperatures and central cooling times, measured with
  $\chandra$, and information about a central radio source from Mittal
  and collaborators.  We determine best-fit relations for different
  subsamples using the cool-core strength and the presence of central
  radio activity as selection criteria. We find the strong cool-core
  clusters~(SCCs) with short cooling times ($<1$~Gyr) to display the
  steepest relation ($\lx \propto \tvir^{3.33\pm0.15}$) and the
  non-cool-core clusters~(NCCs) with long cooling times ($>7.7$~Gyr)
  to display the shallowest ($\lx \propto \tvir^{2.42\pm0.21}$). This
  has the simple implication that on the high-mass scale ($\tvir >
  2.5~$keV) the steepening of the $\lt$ relation is mainly due to the
  cooling of the intracluster medium gas. We propose that ICM cooling
  and AGN heating are both important in shaping the $\lt$ relation but
  on different length-scales. While our study indicates that ICM
  cooling dominates on cluster scales ($\tvir>2.5$~keV), we speculate
  that AGN heating dominates the scaling relation in poor clusters and
  groups ($\tvir < 2.5$~keV).

  \hspace*{1cm} The intrinsic scatter about the $\lt$ relation in
  X-ray luminosity for the whole sample is $45.4\%$ and varies from a
  minimum of $34.8\%$ for weak cool-core clusters to a maximum of
  $59.4\%$ for clusters with no central radio source. The scatter does
  not decrease if SCC clusters are excluded from the full sample. We
  find that the contribution of core luminosities within the cooling
  radius $\rc$, where the cooling time is $7.7$~Gyr and gas cooling
  may be important, to the total X-ray luminosities amounts to $44\%$
  and $15\%$ for the SCC and WCC clusters, respectively. We find that
  after excising the cooling region, the scatter in the $\lt$ relation
  drops from $45.4\%$ to $39.1\%$, implying that the cooling region
  contributes $\sim 27\%$ to the overall scatter. The remaining
  scatter is largely due to the NCCs.

  \hspace*{1cm} Lastly, the statistical completeness of the sample
  allows us to quantify and correct for selection effects individually
  for the subsamples. We find the true SCC fraction to be 25\% lower
  than the observed one and the true normalizations of the $\lt$
  relations to be lower by $12\%$, $7\%$, and $17\%$ for SCC, WCC, and
  NCC clusters, respectively.  }

\authorrunning{Rupal Mittal~et~al.}  \titlerunning{ICM cooling in the
  $\hiflux$ sample of Galaxy Clusters}
\maketitle

\section{Introduction}
\label{intro}

Scaling relations in galaxy clusters and groups are of great interest
for the determination of cosmological key parameters
\cite[e.g.][]{Borgani2001,Rosati2002,Reiprich2002,Stanek2006,Reiprich2006b,Mantz2008,Vikhlinin2009,Leau2010}. Of
prime importance is to keep track of the systematics that enter the
slope and normalization determination due to different physical
mechanisms. These physical processes may, for example, be related to
AGN activity, sloshing or bulk motions of gas, or cooling of the
intracluster medium~(ICM).

In the context of cooling flows and AGN heating, there is still a
considerable amount of debate about the X-ray
luminosity~($\lx$)-temperature~($\tvir$) relation
\citep[e.g.][]{Markevitch1998,Voit2001,Fabian1994a,Bower2008,Maglio2007,McCarthy2004}. The
$\lt$ relation currently faces two main challenges. Firstly, the slope
as determined from observations is much higher \citep[$\lx \propto
\tvir^{2.5-3.0}$, ][]{Allen1998,Arnaud1999} than that predicted based
on the self-similarity of halos \citep[$\lx \propto \tvir^{2.0}$,
][]{Kaiser1986,Eke1998}. Secondly, the intrinsic dispersion~(excluding
statistical errors) in the relation is large, which diminishes its
utility in constraining cosmological parameters. As per the results of
\cite{Pratt2009}, the raw scatter about the $\lt$ relation is about
70\% in $\lx$, which decreases significantly on excising the cluster
central regions~(see Section~\ref{lcf}). Both these findings
illustrate our lack of understanding of the physics that governs the
formation and evolution of the largest virialized structures in the
Universe. As a first step towards using galaxy clusters as
cosmological tools, we need to address variations in their physical
and structural properties caused by baryon processes, such as
radiative cooling, heating due to AGN or conduction, and eradicate
their effects on the scaling relations.

\begin{figure*}[t!]
  \hspace*{-0.75cm}
\begin{minipage}{0.499\textwidth}
    \centering
    \includegraphics[width=1.1\textwidth]{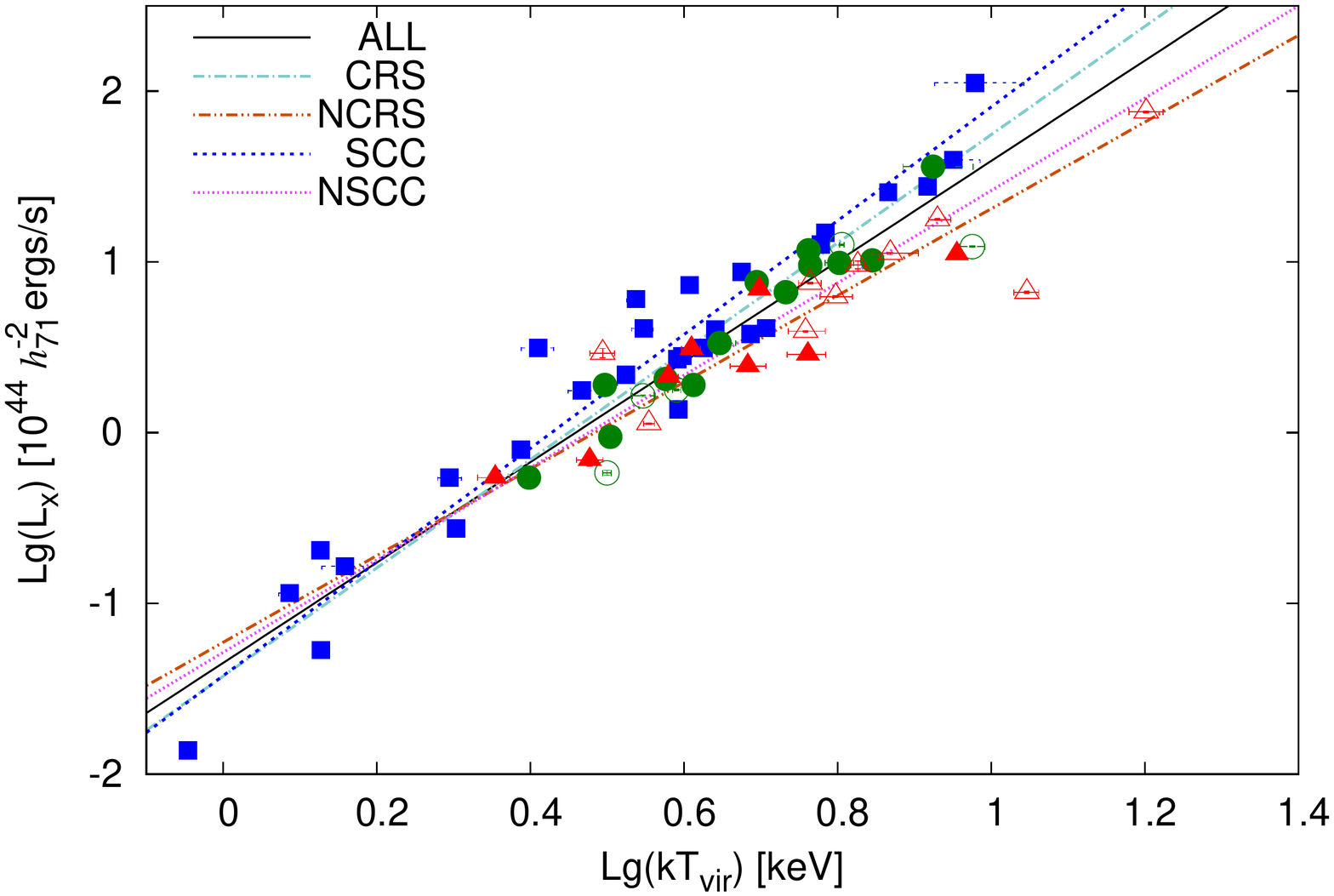}
  \end{minipage}
\begin{minipage}{0.499\textwidth}
    \centering
    \includegraphics[width=1.1\textwidth]{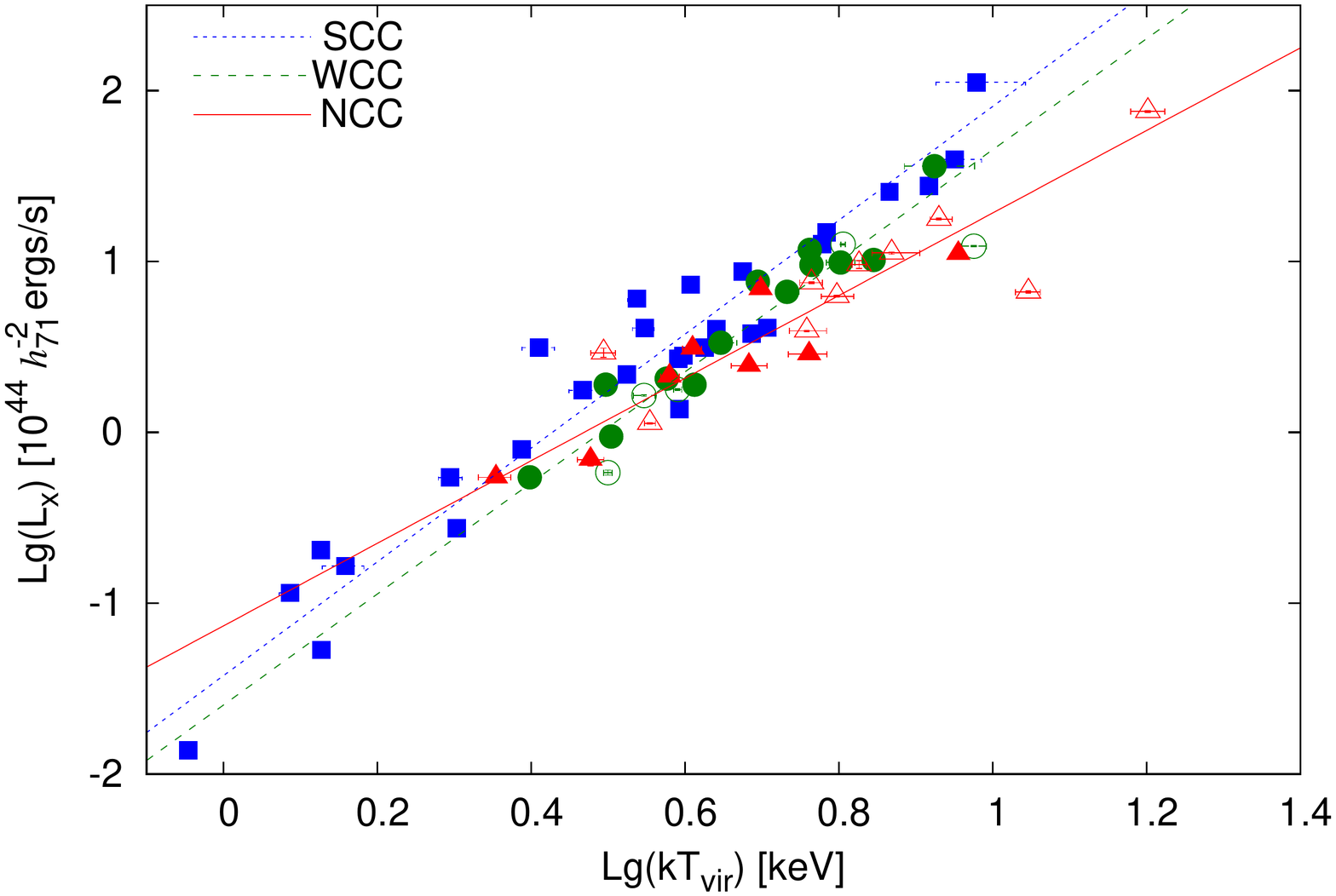}
  \end{minipage}
  \caption{\small $\lt$ relation for the $\hiflux$ sample of clusters.
    The filled symbols represent clusters with a central radio source
    and the open symbols represent those without. The squares (blue),
    circles (green), and triangles (red) represent strong cool-core
    clusters ($\ct <1$~Gyr), weak cool-core clusters ($1$~Gyr $< \ct
    <7.7$~Gyr), and non-cool-core clusters ($\ct >7.7$~Gyr),
    respectively.}
  \label{LT}    
\end{figure*}

Several works, based on both observations and simulations, have
indicated that the $\lt$ relation in low-mass halos with a central
radio source differs systematically from that in systems without a
central radio source
\cite[][]{Sijacki2006,Croston2005,Puchwein2008}. Studies also show
that clusters with strong cooling, as indicated by either the short
cooling times of the intracluster gas or high classical gas mass
deposition rates, lie on a different $\lt$ plane \citep[e.g.,
][]{Allen1998,Hara2006}. In this work, we extend these studies to
understand the behaviour of the $\lt$ relation in the presence of both
AGN heating and ICM cooling. The sample we use to do this is the
complete flux-limited $\hiflux$ sample \citep[][]{Reiprich2002},
constituting the 64 X-ray brightest galaxy clusters with high-quality
$\chandra$ data \citep[][]{PaperIII} and radio data with good spectral
coverage \citep[][]{Mittal2009} for all the clusters. Thus, a rich
database of X-ray and radio data allows us to study the influence of
both the mechanisms in parallel.

\begin{figure*}[t!]
  \begin{minipage}{0.045\textwidth}
    \rotatebox{90}{\Large Residuals}
  \end{minipage}%
  \begin{minipage}{0.95\textwidth}
    \centering
    \hspace*{-2cm}
    \includegraphics[height=0.6\textheight]{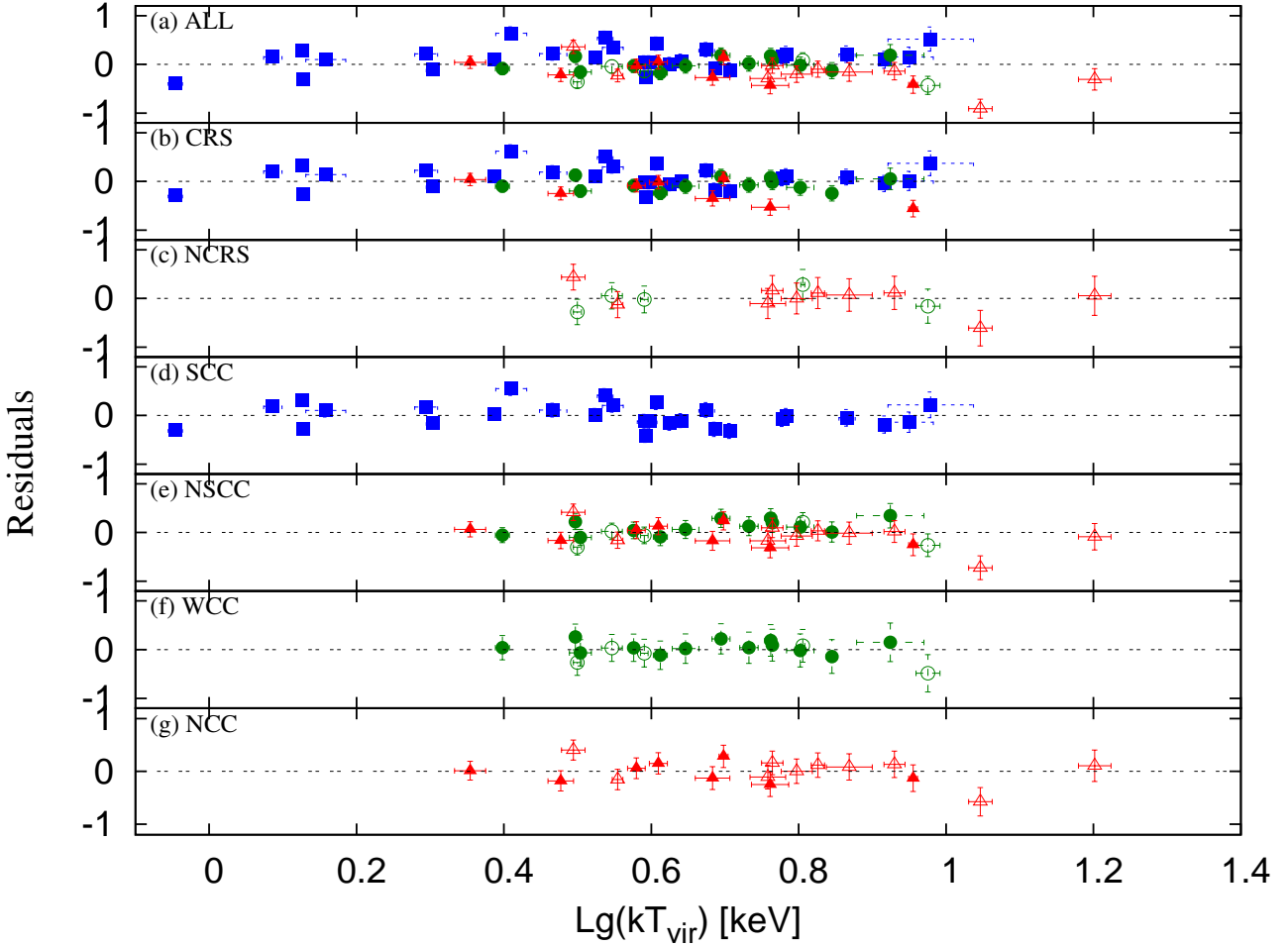}
  \end{minipage}
  \caption{\small The $\lg$-space residuals in $\lx$ for each category
    with respect to their individual best fits. The symbol
    representation is the same as in Fig.~\ref{LT}. The errorbars for
    each subgroup are obtained from summing the uncertainties in the
    measured and the fitted luminosities in quadrature. The fitted
    luminosities for a subgroup are obtained through the best-fit
    $\lt$ relation of that subgroup}.
  \label{dev}    
\end{figure*}

This article is organized as follows. In section~\ref{datanalysis}, we
describe the data used for the work and the data analysis. In section
\ref{L-T}, we outline the subsamples created based on the radio and
cooling properties and compare the individual $\lt$ fits. In section
\ref{lcf}, we scrutinize the cooling activity as the main cause of the
scatter about the observed $\lt$ relation. In section~\ref{sys}, we
investigate systematics, such as the selection effects that might bias
the observed fractions of the different type of cool-core clusters and
whether the new Chandra calibration has any impact on the slope of the
fitted $\lt$ relations. Finally, in sections \ref{disc} and
\ref{conc}, we discuss our results and present our conclusions.  We
assume throughout this paper a $\Lambda$CDM concordance Universe, with
$H_0 = 71~h_{71}$~km/s/Mpc, $\Omega_{\st m} = 0.27$, and
$\Omega_{\Lambda} = 0.73$.

\section{Data and analysis}
\label{datanalysis}

In this section, we present a brief introduction of the datasets
and quantities used in this work. The data reduction and analysis can
be found in detail in \cite{PaperIII}~[hereafter H10] for the X-ray
data and \cite{Mittal2009}~[hereafter M09] for the radio data.

The study in H10 used $>4.5$~Ms of high-resolution $\chandra$ data and
provided a detailed census of the inner regions of clusters offering
an insight into the physics governing cluster cores. It summarized 16
different cool-core diagnostics and found that the central cooling
time, $\ct$, was the best indicator. This choice resulted in 44\%
strong cool-cores (SCC) with $\ct < 1~\hinv{-1/2}$~Gyr, 28\% weak
cool-cores (WCC) with $1~\hinv{-1/2}$~Gyr $<\ct < 7.7~\hinv{-1/2}$~Gyr
and another 28\% non-cool-cores (NCC) with $\ct >
7.7~\hinv{-1/2}$~Gyr. Following this result, we use $\ct$ as the
measure of the cooling strength of a cluster, such that those with
shorter $\ct$ have higher cooling strengths.

M09 utilized more than $140$ different radio flux-density measurements
in their study and obtained the total radio luminosities for
correlation with cooling properties.  They also determined the
fraction of central radio sources~(CRSs) in the $\hiflux$ sample and
found that 48 out of 64 clusters (75\%) contain cluster-centre radio
sources that are either cospatial with or within 50$\hinv{-1}$kpc of
the X-ray peak emission. In addition, M09 found that the probability
of finding a cluster-centre radio source increases with cooling
strength, from 45\% to 67\%\footnote{In the meantime, one more WCC
  cluster~(A1650) has been found, owing to higher sensitivity of the
  more recent radio observations, to harbour a central radio source
  \citep[][]{Govoni2009}, increasing the total number of
  cluster-centre radio sources from 48 to 49 and the fraction of CRSs
  in the WCC category from 67\% to 72\%.} to 100\% for NCC, WCC and
SCC clusters, respectively. This provides evidence of a connection
between the supermassive black hole activity and the cooling of the
ICM.

The virial temperatures, $\tvir$, were taken from H10. The authors
determined $\tvir$ by removing the central region of the observed
$\chandra$ temperature profile. This was done in order to prevent the
cool-core region from corrupting the global cluster temperature
estimates. To determine the size of the central region to be excluded,
we fitted the temperature profiles to a broken power-law. The radius
of the central region in the power-law was free and the index of the
outer component was fixed to be zero.  Owing to the limited
field-of-view of $\chandra$, the outer radius for the $\tvir$
determination varies. This may introduce a small amount of scatter,
although, tests on two of the highest-redshift clusters in the sample,
A2204 and A2163, indicate that our conclusions are not affected by
this.

The bolometric X-ray cluster luminosities over the range (0.01 to
40)~keV, $\lx$, as measured with {$\rosat$} using mostly ASCA
temperatures, were taken from \cite{Reiprich2002}.  $\rosat$ data were
preferred over those of $\chandra$ for the total cluster luminosities
because $\rosat$ has a larger field of view and a lower instrumental
background than $\chandra$.  The errors in the bolometric luminosities
are reflective of the errors in the count rate. Temperature
uncertainties, which are on the order of 10\% \citep{Reiprich2002},
form an additional source of statistical scatter. On the basis of on
simulations using PIMMS~(portable, interactive multi-mission
simulator) and theoretical grounds ($\lx \sim \tvir^{0.5}$), a 10\%
error in temperature implies a $\sim 5\%$ error in $\lx$. Although we
do not take this additional uncertainty into account, one may
conservatively add a 5\% error to the statistical uncertainty in
quadrature. To scrutinize the cooling effects on the $\lt$ relation,
we used, as the first step, the total X-ray luminosities,
i.~e. without performing any correction to avoid the contribution from
the cool-core region. As the second step, we made the cool-core
correction to the luminosities by excluding the contribution to the
luminosity from a central region where cooling is putatively
important. The size of the excluded central region was determined
using the ``cooling radius'', $\rc$, defined as the radius at which
the gas cooling time is $7.7~$Gyr~(the cooling time at radii greater
than $\rc$ is longer than this time~scale). For comparison, we also
considered excluding a region of radius a fixed fraction of the virial
radius, known as the ``core radius'', $\rcore$.

Luminosities within the cooling radius, $\lxrc$, were determined using
CIAO 4.2 and CALDB 4.3.0. and the CIAO tool {\it specextract}. For
each cluster with $\rc > 0$, spectra were extracted from within the
cooling radius for each ObsID in which the entire aperture fit on a
single CCD.  Blank sky backgrounds and weighted response functions
were used, and each spectrum was binned by 25 counts per bin.  Spectra
for each cluster were simultaneously fit in XSPEC using a wabs*APEC
spectral model over the range (0.3 to 10.0)~keV, and allowing the
temperature, abundance, and normalization to vary. The hydrogen
absorbing column densities were fixed to the values used for the H10
study, which were in turn taken, except in a few cases, from the
Leiden/Argentine/Bonn H{\sc i} Survey \citep{Kalberla2005}. The
best-fitting model parameters were used with the XSPEC {\it
  'dummyrsp'} command to provide unabsorbed $\lxrc$ within the energy
range (0.01 to 40)~keV.

For four of the clusters (A0262, NGC4636, A3526, and NGC5044), the
cooling radius proved to be larger than a single chip. In those cases,
spectra were extracted from neighboring chips and fit simultaneously,
tying the temperatures and abundances together but allowing the
normalizations to vary separately. When neighboring chips were too
small to contain a complete annulus, Lx within the total annulus was
estimated using simple geometric arguments.  Note that $\tvir$ and
$\lxrc$ were taken from H10, wherein the X-ray data acquired with
$\chandra$ were calibrated and analysed using CIAO~3.2.2 and
CALDB~3.0. To confirm consistency between the different versions of
CIAO and CALDB, we compared values of $\lxc$ and $\lxrc$ in cases
where $0.048\rvir$ $\approx \rc$ and found that they agreed to within
their $1\sigma$ errors.  The various X-ray quantities used for this
paper are listed in Table~\ref{obs}.

\begin{table*}[t!]
  \centering
  \caption{The best-fit bolometric $\lt$ relation, given by $\frac{\lx}{10^{44}~h_{71}^{-2}~{\st{erg/s}}} = \alpha~\times~\left(\frac{\tvir}{4~{\st{keV}}}\right)^{\beta}
    $, for the individual subcategories of the $\hiflux$ sample of galaxy clusters. 
    \comment{Also given are the intrinsic and statistical scatters in
      percentages, $\sigma_{\st{raw}}$ and $\sigma_{\st{stat}}$ (defined
      in Section~\ref{L-T}), in both the variables, $\lx$ and $\tvir$}} 
  \label{best-fit}
  \begin{tabular}{|c|c|c|c|c|c|c|c|}
    \hline
    Category & \# & $\alpha$ & $\beta$ 
    & $\sigma_{\st{int,~\lx}}$ (in \%)  & $\sigma_{\st{stat,~\lx}}$ (in \%) 
    & $\sigma_{\st{int,~\tvir}}$ (in \%) & $\sigma_{\st{stat,~\tvir}}$ (in \%) \\
    \hline\hline
    (a) & \multicolumn{7}{|c|}{Total luminosities } \\
    \hline
    ALL    & 64 & 2.64$\pm$0.20 & 2.94$\pm$0.16 & \comment{46.9} 45.4 & 11.8 & \comment{16.0} 15.5 & 4.0 \\
    CRS    & 49 & 3.04$\pm$0.23 & 3.17$\pm$0.15 & \comment{48.2} 46.5 & 12.7 & \comment{15.2} 14.6 & 4.0 \\
    NCRS & 15 & 2.00$\pm$0.35 & 2.54$\pm$0.27 & \comment{60.3} 59.4 &  10.3 & \comment{23.8} 23.4 & 4.0 \\
    SCC   & 28 & 3.82$\pm$0.38 & 3.33$\pm$0.15 & \comment{53.4} 51.8 & 13.1 & \comment{16.1} 15.6 & 3.9 \\
    $^{*}$NSCC & 36 & 2.19$\pm$0.17 & 2.70$\pm$0.19 & \comment{48.7} 47.4 & 11.1 & \comment{18.0} 17.5 & 4.1 \\
    WCC  & 18 & 2.30$\pm$0.20 & 3.25$\pm$0.32 & \comment{36.9} 34.8 & 12.3 & \comment{11.4} 10.7 & 3.8 \\
    NCC  & 18 & 2.10$\pm$0.26 & 2.42$\pm$0.21 & \comment{51.5} 50.4 & 10.8 & \comment{21.3} 20.8 & 4.5 \\
    \hline \hline
    (b) & \multicolumn{7}{|c|}{Cool-core corrected luminosities (using the cooling radius, $\rc$, see Section~\ref{lcf}) } \\
    \hline
    ALL    & 64 & 1.88$\pm$0.11 & 3.05$\pm$0.14 & \comment{46.9} 39.1 & 12.5 & \comment{16.0} 12.8 & 4.1 \\
    SCC    & 28 & 2.00$\pm$0.16 & 3.27$\pm$0.15 & \comment{46.9} 34.2 & 13.6 & \comment{16.0} 10.4 & 4.2 \\
    $^{*}$NSCC    & 36 & 2.01$\pm$0.15 & 2.70$\pm$0.18 & \comment{47.8} 47.9 & 11.2 & \comment{16.0} 17.7 & 4.1 \\
    WCC    & 18 & 1.95$\pm$0.15 & 3.20$\pm$0.29 & \comment{46.9} 36.3 & 12.4 & \comment{16.0} 11.4 & 3.9 \\
    $^{*}$CC    & 46 & 1.97$\pm$0.11 & 3.25$\pm$0.13 & \comment{46.9} 33.2 & 13.0 & \comment{16.0} 10.2 & 4.0 \\
    \hline
    \multicolumn{8}{c}{$^{*}$ CC = SCC+WCC, NSCC = WCC+NCC } \\
  \end{tabular}
\end{table*}

\section{$\lt$ scaling relation}
\label{L-T}

The main purpose of this investigation is to assess the effects of
different physical mechanisms on the ICM properties. In particular, we
wish to distinguish the roles of cooling activity and AGN heating on
the $\lt$ relation. Shown in Figure~\ref{LT} is $\lx$ versus
$\tvir$. At first glance, it appears that clusters with a central
radio source (filled symbols) have a systematically higher X-ray
luminosity than those without (open symbols). This would be rather
surprising since AGN heating is believed to heat the X-ray emitting
gas resulting in an opposite effect, namely, an increase in entropy
and suppression of the X-ray luminosity. A more thorough analysis
presented below reveals that it is the cooling of the intracluster gas
in the cores of clusters, many of which harbour an AGN at their
centres, that results in an enhancement in the X-ray luminosity.

To examine the effects of AGN heating and ICM cooling on the $\lt$
relation and to what extent these mechanisms are responsible for the
steepening of the relation, we divided our sample into seven
categories:~(1)~all clusters~(ALL), (2)~clusters with a central radio
source~(CRS), (3)~clusters without a central radio source~(NCRS),
(4)~SCC clusters, (5)~non-strong-cool-core~(NSCC=WCC+NCC) clusters,
(6)~WCC clusters, and (7)~NCC clusters. We used the BCES (bivariate
correlated errors and intrinsic scatter) fitting routine by
\cite{Akritas1996} to determine the best-fit relations for all the
categories, individually. The BCES algorithm generates four different
kinds of fits, amongst which we use the `bisector' method throughout
this paper. This produces a line that bisects the best-fit regression
lines, BCES($X1/X2$) and BCES($X2/X1$), where BCES($X1/X2$) minimizes
the residuals in $X1$ and BCES($X2/X1$) minimizes the residuals in
$X2$. The bisector method ensures that both the quantities, $\lx$ and
$\tvir$, are treated symmetrically, without having to specify
independent~(explanatory) and dependent~(response) variables. The
best-fit lines are shown in Figure~\ref{LT}. We use a power-law
functional form to indicate the $\lt$ relation
\mbox{[$\frac{\lx}{10^{44}~h_{71}^{-2}~{\st{erg/s}}} =
  \alpha~\times~\left(\frac{\tvir}{4~{\st{keV}}}\right)^{\beta} $]}
and perform the fittings in the \mbox{$\ln$-space}. The best-fit model
parameters are given in Table~\ref{best-fit}(a).

The raw scatter~(statistical plus intrinsic) was determined from the
weighted sample variance in the $\lg-\lg$ plane\footnote{$\lg(x) =
  \log_{10}(x)$ and $\ln(x) = \log_{\st e}(x)$}
\citep[][]{Arnaud2005}

\begin{eqnarray}
  \sigma_{\st{raw,~\lx}}^2 & = & {\st C_{\lx}} \sum_{i=1}^N \frac{1}{\sigma_{i,~\lx}^2}\left[Y_i - (\lg\alpha + \beta X_i) \right]^2 \nonumber \\
  {\st C_{\lx}} & = & \frac{1}{(N-2)} \frac{N}{\sum_{i=1}^N (1/\sigma_{i,~\lx}^2)} \, \\
  \sigma_{\st{raw,~\tvir}}^2 & = & {\st C_{\tvir}} \sum_{i=1}^N \frac{1}{\sigma_{i,~\tvir}^2}\left[X_i - (Y_i - \lg\alpha)/\beta \right]^2 \nonumber \\
  {\st C_{\tvir}} & = & \frac{1}{(N-2)} \frac{N}{\sum_{i=1}^N (1/\sigma_{i,~\tvir}^2)} \, ,
\end{eqnarray}
where $Y_i = \lg({\lx}_i)$, $X_i = \lg({\tvir}_{,\,i})$,
${\sigma}_{i,~\lx}^2 = (\Delta Y_i)^2 + \beta^2 (\Delta X_i)^2 $,
${\sigma}_{i,~\tvir}^2 = (\Delta X_i)^2 + (\Delta Y_i)^2/\beta^2 $,
and $N$ is the sample size. The statistical scatter,
$\sigma_{\st{stat}}$, caused by measurement errors, was estimated by
taking the root-mean-square of $\sigma_i$. The intrinsic scatter was
calculated as the difference between the raw and the statistical
scatters in quadrature. In Table~\ref{best-fit}(a), we provide the
intrinsic and statistical scatters in percentages
(\mbox{$\sigma*\ln(10)*100$}) in both, $\lx$ as well as $\tvir$. Since
the statistical and intrinsic scatters add in quadrature, the
contribution of the statistical errors to the total dispersion is
small.

\subsection{Cooling against heating}
\label{coolheat}

The best-fit slope of the luminosity-temperature relation for the
complete $\hiflux$ sample~(category `ALL') is \mbox{$2.94\pm0.16$},
which is much steeper than the self-similar value of 2.0. This result
agrees with several previous works such as \cite{Arnaud1999},
\cite{Allen1998}, \cite{Novicki2002}, and \cite{Zhang2007}.  After the
subdivision into different classes, several interesting features are
evident. A comparison of slopes shows that the SCC clusters have the
steepest relation ($\beta=3.33\pm0.15$), whereas the NCC clusters have
the shallowest ($\beta=2.42\pm0.21$). Note that the high power-law
index for SCCs is not due to selection effects. Even though it is true
that there is not even a single NCC with $\tvir<2.5$~keV, the steep
$\lx-\tvir$ relation for SCCs is a generic feature of clusters with a
short central cooling time. This can be easily verified by eliminating
from the SCC subsample all clusters with $\tvir<2.5$~keV~(eight in
number) and fitting the data again. After doing this, we find that
$\beta=3.14\pm0.34$, which is consistent with the power-law index
obtained for the complete SCC sample, and still much higher than the
fitted power-law index for the NCC subsample.

A similar contrast in slopes is true for the CRS and NCRS
clusters. The fits for CRS and SCC clusters have consistent slopes
within the errorbars, and, likewise, the fits for NCRS and NCC
clusters have consistent slopes within the errorbars (although the CRS
and SCC cluster subsamples have different normalizations, the SCC
clusters being systematically higher in luminosity than the CRS
clusters). This is not unsurprising since we know from M09 that the
probability of finding a central radio source in a cluster is an
increasing function of the cooling strength, with SCC clusters having
the highest incidence~(100\%) and the NCC clusters having the lowest
incidence~(44\%). Hence, the fit through the CRS cluster subsample is
dominated by the SCC clusters and, similarly, the fit through the NCRS
cluster subsample is dominated by the NCC clusters.

In Figure~\ref{dev}, we present the $\lg$-space residuals in $\lx$ for
each category based on its respective best-fit. Figure~\ref{dev}(b)
indicates that the CRS subsample actually comprises clusters from two
different populations rather than one. CRS clusters with a strong cool
core (filled boxes) are clearly above the zero-deviation line (18 out
of 28), whereas those without (circles and triangles) are on-average
below (15 out of 21). Using the Kolmogorov-Smirnov~(K$-$S) test, we
find that the probability of the null hypothesis, i.e. that the two
samples (CRS clusters with and without a SCC) are drawn from the same
underlying population, is only 1.2\%.  A similar, though less
dramatic, segregation is seen between NSCC clusters with a CRS,
represented by the filled symbols in Figure~\ref{dev}(e) and those
without a CRS, represented by the open symbols in
Figure~\ref{dev}(e). Fourteen of the 21 NSCC clusters with a CRS lie
above the zero-deviation line and 6 out of 15 NSCC clusters without a
CRS lie above the zero-deviation line. The K$-$S probability of the
null hypothesis is 7\%. NSCC clusters with a CRS are still biased
towards positive deviation. This is likely because a higher fraction
of WCC clusters have a CRS than the NCC, once again implying that the
cooling activity dictates the overall trend. The NCC category in
Figure~\ref{dev}(g), even though low in sample size, most clearly
illustrates this argument. Both the subsamples, the NCC clusters with
(filled symbols) and without (open symbols) a CRS, can be seen equally
distributed about the zero-deviation line. Four of the 8 NCC clusters
with a CRS lie above the zero-deviation line and 7 out of 10 NCC
clusters without a CRS lie above the zero-deviation line. The K$-$S
probability that the NCC clusters with and without a CRS originate
from a single population is 54\%. In other words, subsamples marked by
the absence (or presence) of cooling may be visualized as originating
from a single population. That the subsamples having identical cooling
properties seem to follow the same $\lt$ relation, independent of the
presence of a radio source, implies that it is the cooling of the
intracluster gas that plays a more dominant role in influencing the
$\lt$ relation rather than the AGN activity.

We conclude that on the high-mass scale ($T_{\st vir} > 2.5$~keV) the
steepening of the $\lt$ relation is primarily related to the cooling
of the ICM. This is clearly manifested by the $\lt$ relation being the
steepest for the SCC clusters. As the forthcoming analysis shows, the
cooling effects in CCs leading to the steeper relation are likely
distributed over a region larger than that described by the cooling
radius.  The cut at $\tvir=2.5$~keV is motivated by the SCC and NCC
fits crossing at $\lg(k\tvir) \sim 0.4$ (see Figure~\ref{LT}). As
discussed in Section~\ref{disc}, it is most likely that both ICM
cooling and AGN heating have to be considered in describing the $\lt$
relation.  However, owing to the nature of the sample containing
mostly high-mass clusters, we see direct evidence of only the former.

\section{Cool-core contribution to the X-ray luminosity and scatter
  in the $\lt$ relation}
\label{lcf}

\begin{figure*}
  \hspace*{-0.75cm}
  \begin{minipage}{0.499\textwidth}
    \centering
    \includegraphics[width=1.1\textwidth]{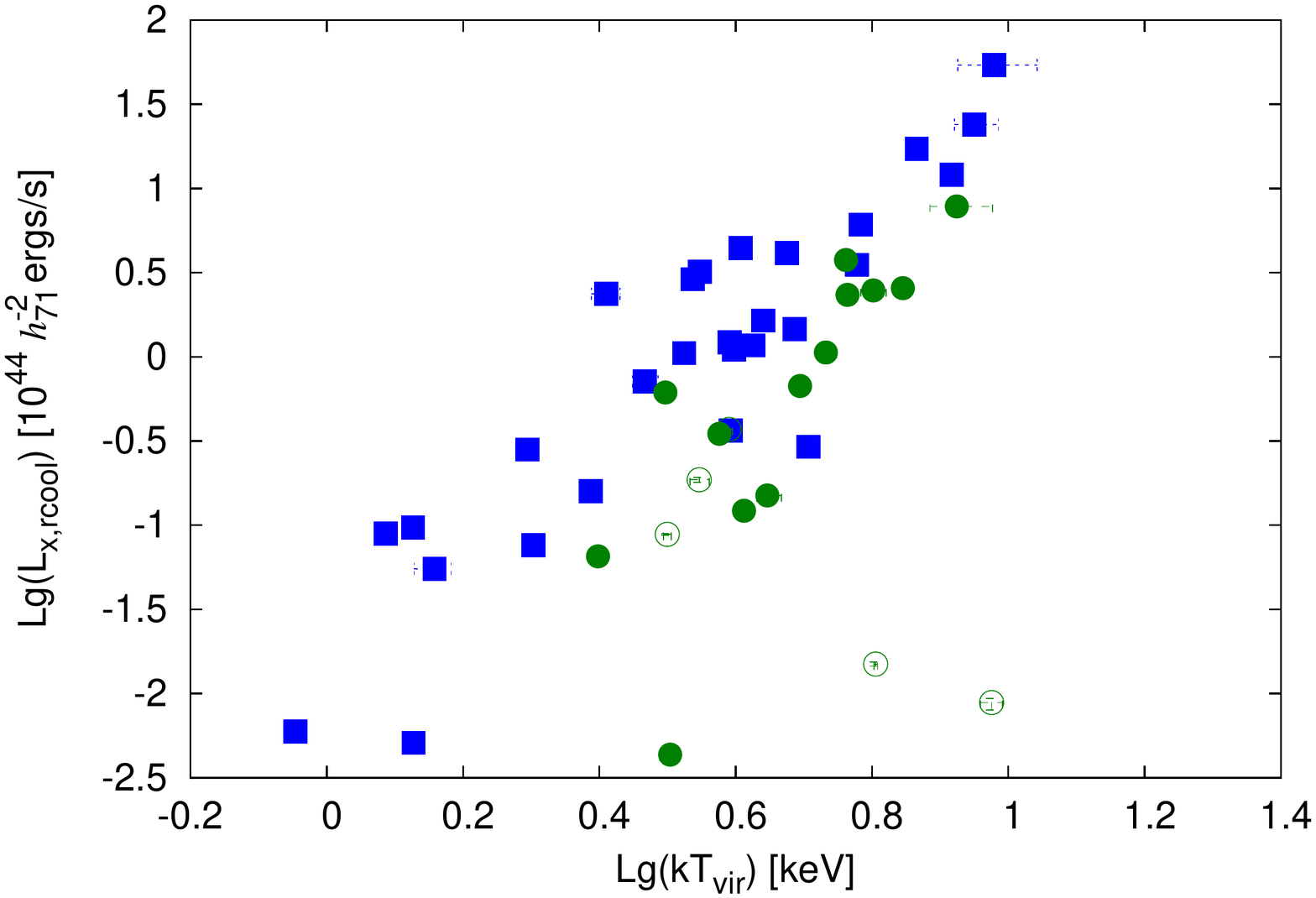}
  \end{minipage}%
  \begin{minipage}{0.499\textwidth}
    \centering
    \includegraphics[width=1.1\textwidth]{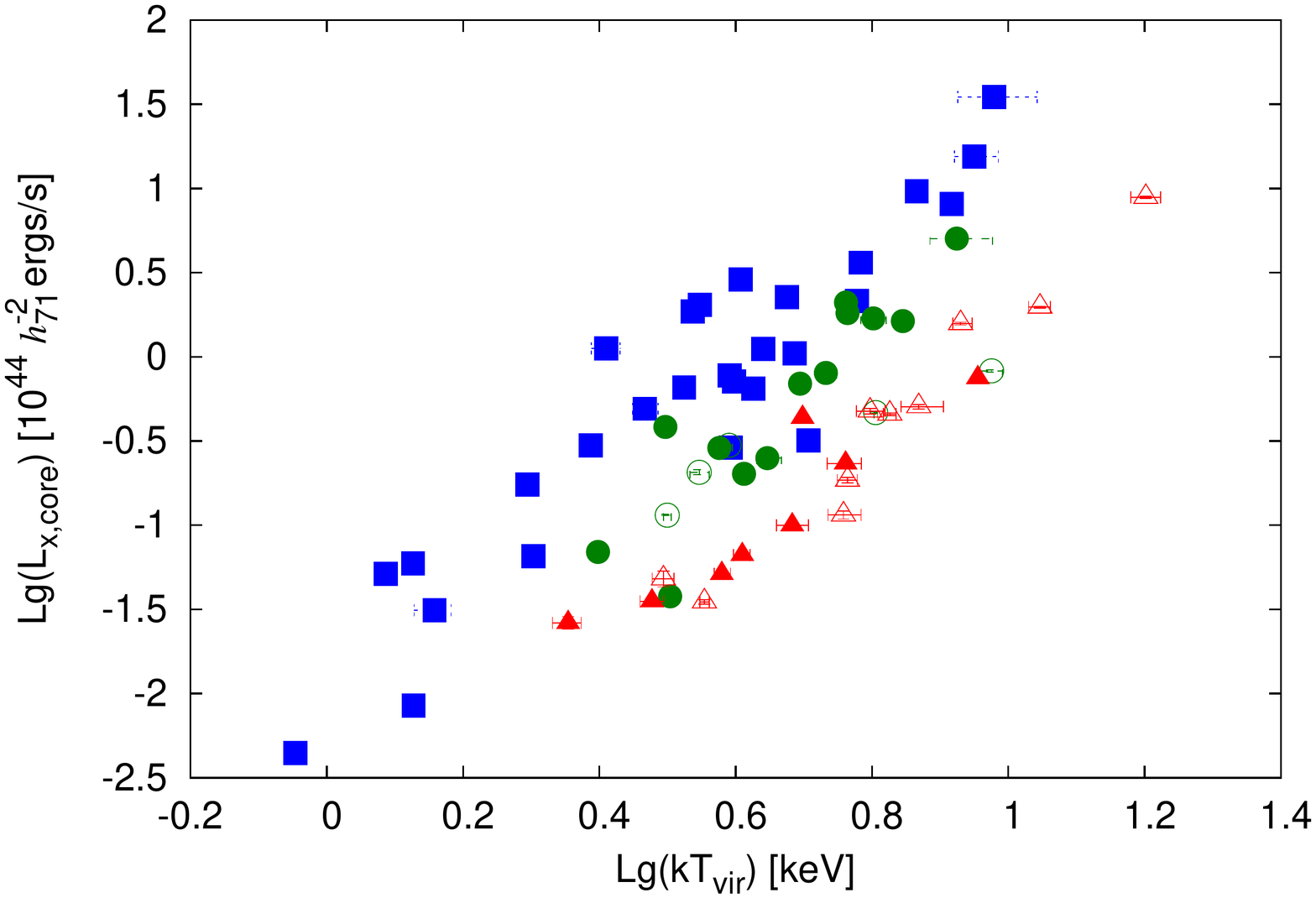}
  \end{minipage}\\
  \hspace*{-0.75cm}
  \begin{minipage}{0.499\textwidth}
    \vspace*{-1cm}
    \centering
    \includegraphics[width=1.1\textwidth]{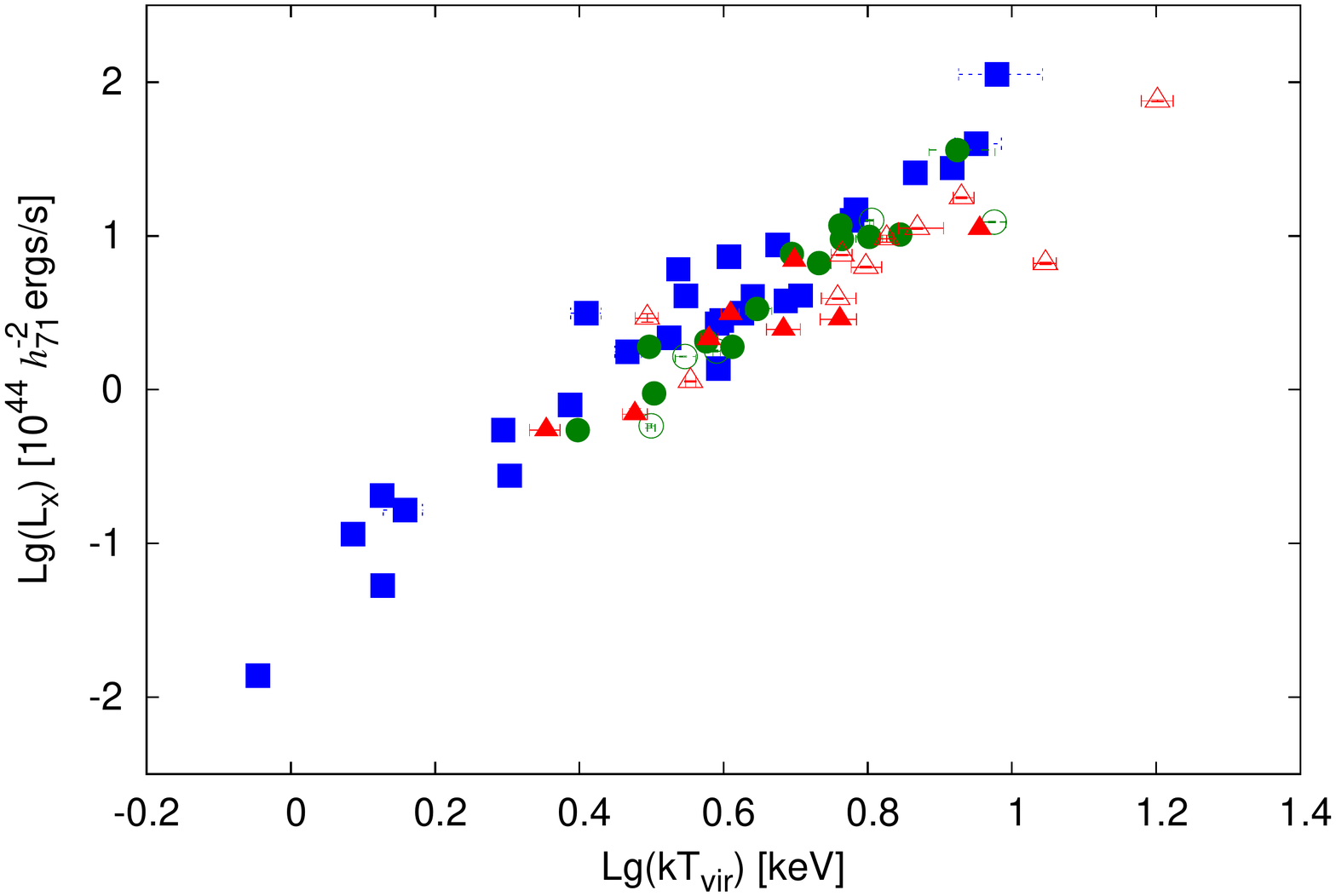}
  \end{minipage}%
  \begin{minipage}{0.499\textwidth}
    \vspace*{-1cm}
    \centering
    \includegraphics[width=1.1\textwidth]{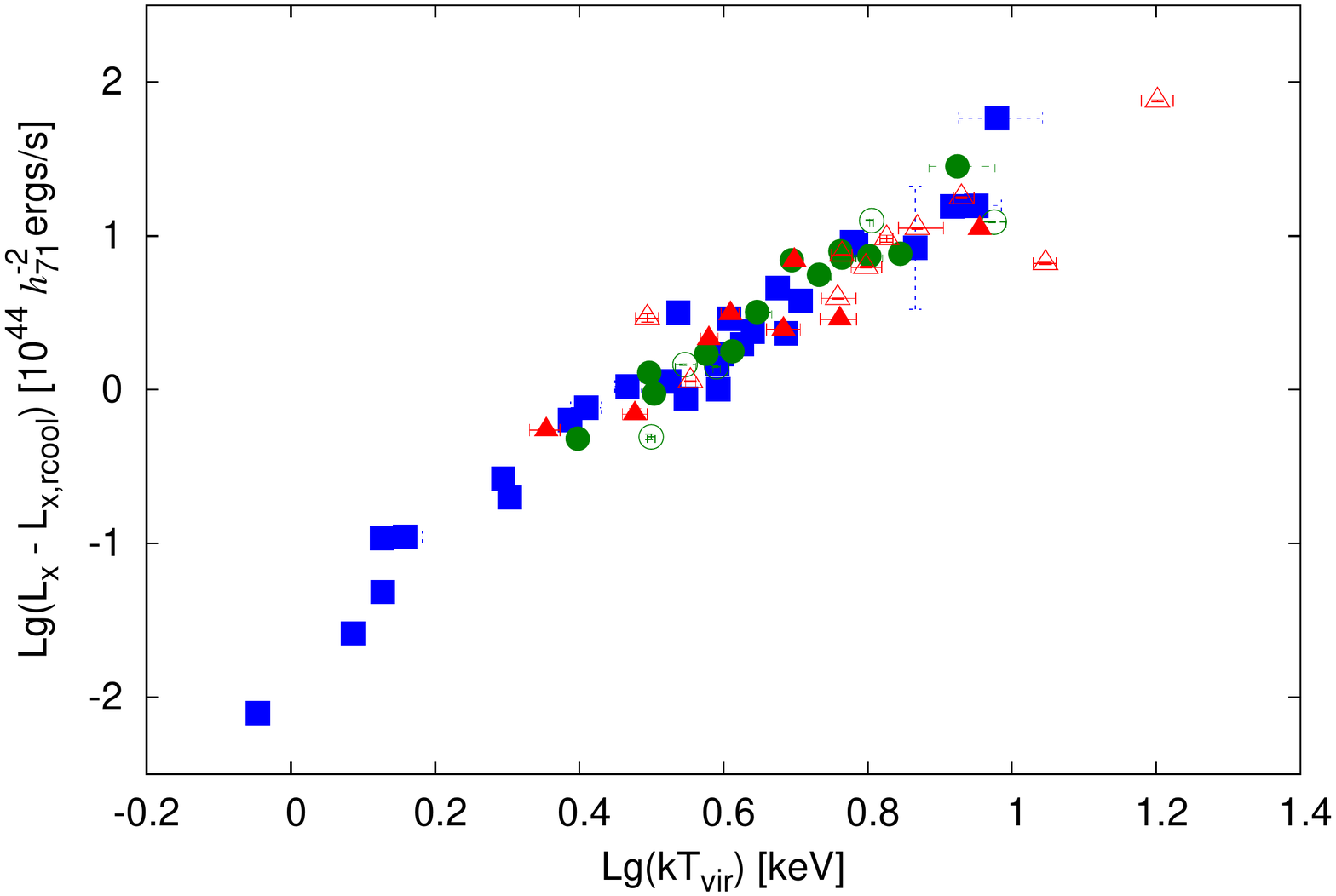}
  \end{minipage}
  \caption{\small {\it Upper panel}: The X-ray luminosity
    within the cooling radius at which the cooling time equals
    7.7~Gyr, $\lxrc$ (left), and that within the core radius of 4.8\%
    $\rvir$, $\lxc$ (right), vs temperature. {\it Lower panel}: The
    total X-ray luminosity vs temperature (left) and the 
    total X-ray luminosity minus $\lxrc$ vs temperature (right). 
    Shown are the three categories $-$ strong cool-core~(SCC), weak
    cool-core~(WCC) and non-cool-core~(NCC) clusters. The symbol
    representation is the same as in Fig.~\ref{LT}.}
  \label{flxc} 
\end{figure*}

It has often been noted that excluding cool-core clusters decreases
the overall intrinsic dispersion in the $\lt$ relation
\citep[][]{Hara2006,Pratt2009,Chen2007}, with the direct implication
that the cool-core related activities are the prime contributors to
the observed scatter. Our observations do not completely support this
conclusion. Qualitatively, this may be seen by noting that, firstly,
if that were true, the intrinsic scatter for the subgroup `NSCC'
containing no strong cool-core clusters would have to be significantly
lower than that corresponding to the `ALL' category.  From
Table~\ref{best-fit}(a) it can be seen this is not the case.
Secondly, the subgroup `NCRS', which comprises predominantly NCC
clusters and not a single SCC cluster has the largest intrinsic
dispersion in both X-ray luminosity~(59.4\%) as well as virial
temperature~(23.4\%). Also, the `NCC' subgroup has a large a high
intrinsic scatter~(50.4~\%), comparable to the `SCC'
subgroup~(51.8\%). That these two subsamples have comparable scatters
implies that one cannot impute the scatter obtained for the 'ALL'
category to only the process of intracluster-medium cooling. In H10,
it was found that most of the NCC clusters in the $\hiflux$ sample
show signs of merger activity, a process that entails shock heating
and adiabatic compression and may cause the system to deviate from a
given $\lt$ relation.

\begin{figure}
  \centering
  \includegraphics[width=0.5\textwidth]{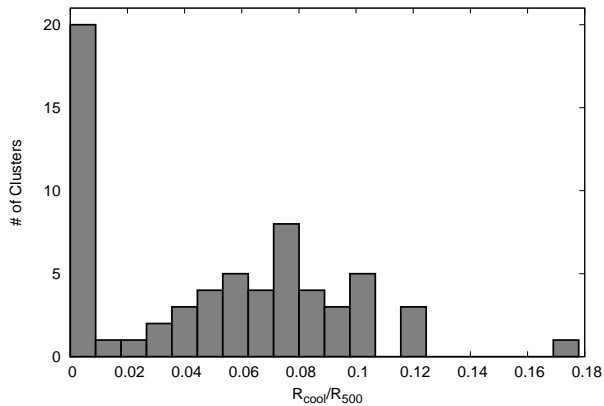}
  \caption{\small A histogram of the scaled cooling radius,
    $\rc/\rvir$, showing a large dispersion. There are 18
    non-cool-core clusters with $\rc=0$ and 46 cool-core clusters
    with an average cooling radius of \mbox{$(0.07\pm0.03)\rvir$}.}
  \label{Rcool-R500}    
\end{figure}

In the following, we quantify the degree to which the cooling activity
contributes to the observed scatter about the $\lt$ relation. To
calculate the X-ray luminosity originating from the cool cores of the
clusters, the ``cool-core luminosity'', we choose two defining
regions. The first region is the ``cooling radius'', $\rc$, the radius
at which the cooling time of the gas is $7.7$~Gyr. The NCCs by
definition have $\rc=0$. The average cooling radius of the remaining
46 cool-core~(CC) clusters (SCCs+WCCs) is $\sim (0.07 \pm 0.03)
\rvir$. The second region is the ``core radius'', $\rcore=0.048 \rvir$
(H10, unpublished), a fixed fraction of the virial radius. The
cool-core luminosities (0.01-40)~keV from $\rc$ and $\rcore$ versus
temperature are shown in the upper left and right panels of
Fig.~\ref{flxc}, respectively. The left panel shows those SCCs and WCCs
with a non-zero $\rc$, whereas the right panel shows all the 64
clusters. Even though $\rcore$ closely matches the average cooling
radius ($\sim 0.05\rvir$) over all 64 clusters, $\rc$ is likely a
better parameter to assess the scatter due to cooling in the $\lt$
relation. $\rcore$ shows a relatively smaller scatter with temperature
than $\rc$. However, this is not surprising. The core radius
corresponds to a fixed fraction of a characteristic overdensity scale,
whereas the cooling radius varies from cluster to cluster and its
scaling relative to the $\rvir$ has a large dispersion (see
Fig.~\ref{Rcool-R500}). Since the cooling radius is closely tied to
the cooling process, it is precisely this dispersion that may be
adding to the scatter in the $\lt$ relation. For samples where
detailed density and temperature profiles cannot be obtained (such as
those containing high-$z$ clusters) for the precise determination of
$\rc$, we propose that a cut equal to the average cooling radius $\sim
0.07 \rvir$ be used for the cool-core correction. A cut larger than
this is likely to overestimate the fraction of the luminosity
generated by cooling.

We calculate the fraction of the X-ray luminosity originating from the
cooling radius as \mbox{$\flxrc = \lxrc/\lx$}, and that originating
from the core radius as \mbox{$\flxc = \lxc/\lx$}, where $\lxrc$ and
$\lxc$ are the integrated luminosities over $\rc$ and $\rcore$,
respectively. We find that the $\flxrc$ for the SCCs is equal to
$44.0\%\pm14.2\%$, for the WCC is equal to $15.2\%\pm15.5\%$, for the
SCCs and WCCs combined is equal to $31.4\%\pm31.6\%$, and for all the
64 clusters is $23.5\%\pm 23.5\%$ (the 1$\sigma$ error in $\flxrc$ is
greater than the mean for some of the subsets and reflects
non-gaussianity in $\flxrc$). Similarly, the $\flxc$ for the whole
sample is equal to $17.8\%\pm13.1\%$. Hence, the variation in total
$\lx$ due to the variation in $\lxrc$ for the whole sample ($23.5\%$)
is much higher than that due to the variation in $\lxc$
($13.1\%$). These variations may be directly compared to the
percentage scatter in $\lx$ determined for the `ALL' $\lt$
relation~($45.4\%$). The contribution of the cooling activity to the
scatter in the $\lt$ relation is $(23.5\%/45.4\%)^2 \sim 27\%$ if the
cooling radius is considered and $(13.1\%/45.4\%)^2 < 9\%$ if the core
radius is considered (note that the contributions to the scatter add
in quadrature). However, as mentioned above, we deem that the region
described by the cooling radius is a more appropriate region to use to
assess the luminosity due to cooling and its contribution to the
scatter. Hence, this implies that we can expect about 25\% of the
scatter in the $\lt$ relation to result from the cluster-to-cluster
variation in the central luminosities of CC clusters. The scatter
expected after excluding the cool-core luminosity is $45.4\%
\sqrt{1.0-(23.5\%/45.4\%)^2} \sim 39\%$.

In the lower left panel of Fig.~\ref{flxc}, we show the total X-ray
luminosity versus temperature and in the lower right panel of
Fig.~\ref{flxc} the cool-core excised total luminosity versus the
virial temperature. We determine the $\lt$ relation after excluding
the cooling luminosity and find
\mbox{$\frac{\lx}{10^{44}~{\st{erg/s}}} =
  (1.88\pm0.11)~\times~\left(\frac{\tvir}{4~{\st{keV}}}\right)^{3.05\pm0.14}
  $}. The exclusion of the core emission does not affect the slope of
the $\lt$ relation and is consistent with that corresponding to the
`ALL' category including the core~($2.94\pm0.16$). However, it does
produce a smaller normalization and a shallower slope for the SCCs,
such that the SCCs and WCCs are now indistinguishable and have the
same $\lt$ parameters. The best-fit $\lt$ parameters for the subsets
after the cool-core correction, i.~e. after subtracting $\lxrc$ from
the total luminosities, are given in Table~\ref{best-fit}(b). Since no
correction was made for the NCCs, the $\lt$ parameters and the
scatters in the luminosity and temperature for the NCC subset remain
the same. The intrinsic dispersion about the $\lt$ relation for `ALL'
clusters obtained after the cool-core correction is 39.1\%. This is in
very good agreement with the scatter expected after excising the
cooling luminosity based on the dispersion in $\flxrc$, as proposed
above. In other words, about 27\% of the scatter in the $\lt$ relation
may be attributed to the luminosity within $\rc$. However, {\it after}
the cool-core correction, the CC (SCC+WCC) subset shows the least
scatter in luminosity (33.2\%), followed by the SCC subset
(34.2\%). Hence, the remaining scatter in the $\lt$ relation after
excising $\rc$ from the SCCs and WCCs is largely due to the NCCs
(50.4\%). 

We expect the scatter to continually decrease with the size of the
excised region since the effects of cooling and heating are expected
to dampen with increasing clustercentric distance. This is consistent
with the findings of \cite{Pratt2009}, who excised the 15\%$\rvir$
region and found a reduction in scatter by more than $50\%$. We do not
conduct the above analysis with 15\%$\rvir$; the reasons are
twofold. Firstly, because of the limited field-of-view of $\chandra$,
more and more emission falls out of the observed field with increasing
radius (especially for the nearby clusters), forcing us to make
assumptions about the missing emission.  This defeats the purpose of
obtaining precise luminosity measurements to minimize the
scatter. Secondly, as seen with the 4.8\%$\rvir$ core region above,
excising a fixed fraction of the virial radius is not always
meaningful. The actual cooling region scaled by $\rvir$ contributing
to the scatter varies from cluster to cluster.

Note that in the AGN-regulated feedback paradigm, AGN heating must
occur over the same scale as the cooling in order to balance it. While
this analysis shows that excluding the cooling region in the cool-core
clusters reduces the scatter in the $\lt$ relation, we cannot comment
on how much of the scatter from the excised region is caused by
cooling activity and how much by AGN heating.  For a given
temperature, not only might the SCC clusters have a higher luminosity
than NCCs, but for a given luminosity either the SCC clusters might
have lower virial temperatures and/or the NCC clusters higher virial
temperatures.  In their simulations, \cite{Burns2008} indeed find an
abundance of cool gas in the region $0.05\rvir$ to $0.3\rvir$,
i.~e. beyond the cores of CC clusters. This could result in an overall
lower virial temperature in SCC clusters. As noted previously
(Section~\ref{lcf}), most NCC clusters in our sample are merger
systems, which due to heating may have a higher virial temperature.

Lastly, we point out that the subgroup of WCC clusters displays the
least scatter~(34.8\%). This is perhaps not so surprising because
these are clusters that neither possess a strong cool-core nor exhibit
any indication of an ongoing or past major merger
\citep[][]{PaperIII}.  We note that the SCCs and WCCs have
statistically consistent slopes and so it may be that WCCs are
actually progenitors of the SCC clusters, whose cool cores would
eventually grow in time to match those of SCC clusters.

\section{Systematics}
\label{sys}

\begin{table*}
  \centering
  \caption{The impact of selection effects on the observed fractions of 
    SCCs, WCCs, and NCCs and the $\lt$ relations.} 
  \label{simulations}
  \begin{tabular}{|p{3cm}|c|c|c|c|c|}
    \hline
    Case                    & Category & Input Fractions & Output Fractions & Input $\lt$ relations & Output $\lt$ relations  \\
                               &                &                          &                             & $\alpha$, $\beta$    & $\alpha$, $\beta$        \\
    \hline\hline
   \multirow{3}{3.0cm}{No scatter, observed $\alpha$, $\beta$ fixed to 3.33}   
                               & SCC         & 0.310                  & 0.432$\pm$0.040 & 3.82, 3.33                & 3.82, 3.33                    \\  
                               & WCC        & 0.335                  & 0.286$\pm$0.038 & 2.30, 3.33                & 2.30, 3.33                    \\  
                               & NCC        & 0.355                  & 0.282$\pm$0.039 & 2.10, 3.33                & 2.10, 3.33                    \\  
    \hline
   \multirow{3}{3.0cm}{No scatter, observed $\alpha$ and $\beta$}
                               & SCC         & 0.340                 & 0.437$\pm$0.040 & 3.82, 3.33               & 3.82, 3.33                    \\  
                               & WCC        & 0.350                 & 0.276$\pm$0.034 & 2.30, 3.25               & 2.30, 3.25                     \\  
                               & NCC        & 0.310                 & 0.287$\pm$0.038 & 2.10, 2.42               & 2.12, 2.42                    \\  
    \hline 
    \multirow{3}{3.0cm}{Scatter included, observed $\alpha$ and $\beta$}
                               & SCC         & 0.340                  & 0.438$\pm$0.038 & 3.82, 3.33                & 4.31$\pm0.08$,
                                                                                                                                                      3.35$\pm0.01$           \\  
                               & WCC        & 0.355                  & 0.281$\pm$0.037 & 2.30, 3.25                & 2.43$\pm0.02$, 
                                                                                                                                                      3.27$\pm0.01$         \\  
                               & NCC        & 0.305                  & 0.281$\pm$0.038 & 2.10, 2.42                & 2.56$\pm0.07$,
                                                                                                                                                      2.51$\pm0.01$           \\  
    \hline
    \multirow{3}{3.0cm}{Scatter included, observed $\beta$}
                               & SCC         & 0.340                  & 0.432$\pm$0.042 & 3.40, 3.33                & 3.85$\pm0.06$, 
                                                                                                                                                      3.35$\pm0.01$           \\  
                               & WCC        & 0.355                  & 0.291$\pm$0.038 & 2.15, 3.25                & 2.30$\pm0.02$, 
                                                                                                                                                      3.27$\pm0.01$           \\  
                               & NCC        & 0.305                  & 0.277$\pm$0.042 & 1.75, 2.42                & 2.13$\pm0.05$,
                                                                                                                                                         2.51$\pm0.01$        \\
   \hline
 \end{tabular}
\end{table*}

\subsection{Selection effects}
\label{LTsim}

Flux-limited samples suffer from the well-known Malmquist bias, namely
that brighter objects have a higher detection rate than fainter
objects. This bias may affect the observed fractions of SCC, WCC, and
NCC clusters, and, in addition, the fitted $\lt$ relations.

Owing to their enhanced central X-ray-emission. strong cool-core
clusters have a higher chance of detection and this may explain the
observed higher fraction of SCC clusters in flux- and
luminosity-limited samples. Since we have a complete flux-limited
sample, we can estimate the magnitude of this bias. We simulated
samples of clusters, which follow the X-ray temperature function~(XTF)
given by $\st{d} N/\st{d} V \sim T^{-3.2}$ \citep{Markevitch1998}, in
the temperature range (0.5-20)~keV and redshift range from 0.001 to
0.25 [the XTF used is an approximation; a more realistic functional
form consists of a power-law and an exponential high temperature
cut-off \citep[e.g.][]{Henry2000,Ikebe2002}]. In H10, it is shown that
SCC, WCC, and NCC clusters come from the same parent redshift
distributions within a $1\sigma$ standard deviation. Hence, we
assigned random redshifts to clusters conforming to the $N \propto
D^3$ law, where $D$ is the proper distance. We assigned clusters to be
SCC, WCC, and NCC clusters according to certain input fractions, which
were varied in the overall process.  The luminosities of the SCC, WCC,
and NCC clusters were calculated using the $\lt$ relation
corresponding to each category as determined in
Section~\ref{coolheat}.

To estimate the effect of applying a flux-limit to a mixed sample of
SCCs, WCCs, and NCCs, we applied the $\hiflux$ flux-limit, $f_{\textrm
  x}~(0.1-2.4)$~keV$ \ge 2 \times 10^{-11}$~erg~s$^{-1}$~cm$^{-2}$, to
the simulated sample and reprojected the fluxes to the $\lt$ plane.
We tried four different cases. In the first (simplest) case, we
assumed that the SCCs, WCCs and NCCs have the same slope (3.33) and
fixed the normalizations to those found from the fits to the data
(Table~\ref{best-fit})(a). In the second case, we fixed the slopes for
SCCs, WCCs and NCCs to the fitted values.  In the third case, we also
inserted the intrinsic scatter [from Table~\ref{best-fit}(a)] about
the fitted $\lt$ relations in X-ray luminosity. In all three cases, we
varied the input fractions to determine the values that best matched
the observed fractions. The addition of scatter about the $\lt$
relations has an effect, as expected, only on the normalizations. The
data points scattered to higher values have a higher likelihood of
falling above the flux-limit, thereby increasing the normalization in
all three cases. In the fourth case, we fixed the fractions of SCC,
WCC, and NCC to the best-fit values from the third case and varied the
normalizations to determine the `true' values, which yielded the
observed normalizations. The results of these simulations are given in
Table~\ref{simulations}. The two main results are that (1)~the SCC
input fraction is on average about $25\%$ lower than the observed
value and (2)~the inclusion of scatter increases the normalization of
all the three $\lt$ relations, the true normalizations being about
$12\%$, $7\%$, and $17\%$ lower than the observed values for SCC, WCC,
and NCC clusters, respectively.

\subsection{$\chandra$ calibration}
\label{chandra-calib}

The virial temperature acquired with $\chandra$ that were used to
determine the overall $\lt$ relation for the $\hiflux$ sample were
calibrated and analysed using CIAO~3.2.2 and CALDB~3.0
\citep[][]{PaperIII}. In the meantime, a new version of the $\chandra$
calibration package, CALDB~4.1.1, has been made available as of
January~2009. In this section, we give a brief comparison between the
results obtained using the old and the new CALDB packages.

The $\chandra-\XMM$ cross calibration project with the International
Astronomical Consortium for High Energy
Calibration~(IACHEC\footnote{http://www.iachec.org/index.html})
resolved a disagreement between the two instruments to a certain
extent. Internal consistency checks showed that the $\chandra-$ACIS
derived temperatures for clusters hotter than $4$~keV were
systematically higher than the $\XMM-$EPIC derived ones. This may
result in an erroneous slope estimate of the $\lt$ relation.

\begin{figure}
  \centering
  \includegraphics[width=0.5\textwidth]{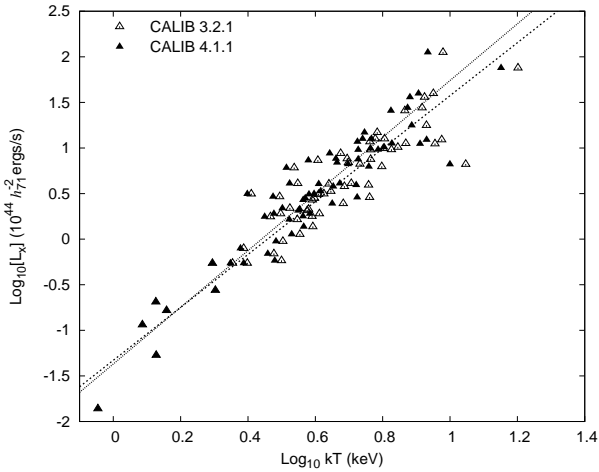}
  \caption{\small The effect of new $\chandra$ calibration
    CALDB~4.1.1. on the $\lt$ relation.}
  \label{caldb}    
\end{figure}

Minor adjustments to the effective area of the HRMA~(High Resolution
Mirror Assembly) have now been incorporated in CALDB~4.1.1, which are
based on the predictions of the new model that corrects for a
hydrocarbon contamination layer. We used the results of the comparison
between cluster temperatures derived for the $\XMM$ and $\chandra$
data, using the two measurements of HRMA effective area \citep[][
figure 21]{David2009}. We determined the best-fit lines to establish
the relation between the ACIS temperatures, $T_{3.2.1}$ and
$T_{4.1.1}$, in the energy band 0.5~keV to 7.0~keV, corresponding to
the two $\chandra$ calibration schemes, CALDB~3.2.1 and
CALDB~4.1.1. This relation is given by
\begin{equation}
T_{4.1.1} = 0.875 * T_{3.2.1} + 0.251 \, ,
\end{equation}
where the temperatures are in units of keV. Using this relation, we
calculated $T_{4.1.1}$ for each cluster with $T_{3.2.1} > 2$~keV. The
old and the new temperature estimates are shown in Figure~\ref{caldb}.
The slope of the best-fit through the new data points (`ALL') obtained
from the BCES algorithm is $3.11 \pm 0.16$, and is consistent within
the errorbars with the previous one given in
Table~\ref{best-fit}(a). Thus, we conclude that our best-fit estimates
of the $\lt$ relation for the different categories obtained using
CALDB~3.1.2 are not affected by the new $\chandra$ calibration.

\section{Discussion}
\label{disc}

The $\lt$ relation in clusters has been studied extensively by several
other groups \citep[e.g.][]{Hara2006,Puchwein2008,Poole2007,
  Jetha2007} to examine the intricate processes operating in the ICM
that need to be well-understood before using X-ray observations of
clusters to constrain cosmological parameters.  Most observational
studies focus either on intracluster cooling or AGN heating as the
cause of the deviation of the $\lt$ relation from self-similar form of
the relation. Our multi-wavelength coverage of the topic allows us to
unravel the effects of both mechanisms, simultaneously.

The general conclusion of previous studies of the AGN effects on the
state of the intracluster gas is that AGN heating removes gas from the
centres of halos
\citep{Sijacki2006,Puchwein2008,Croston2005,Short2009}. For example,
in the model of \cite{Short2009}, the mechanical heating associated
with expanding jets and lobes produced by the central engine with
radio-mode accretion pushes the gas away from the centre to the outer
regions. This has the effect of reducing the density of the X-ray
emitting gas in the halo cores. AGN-heating is especially effective in
poor clusters and groups because of their shallow gravitational
potential wells. In contrast, cooling increases the gas density at the
centre and causes the X-ray luminosity of the cluster cores to
increase. The two processes therefore have opposite effects on the
X-ray luminosity.

These results can be reconciled quite well with our findings. Even
though our observations lend support only to the impact of cooling, we
propose that both the mechanisms have a significant influence on the
$\lt$ relation but on different scales. While we have shown that ICM
cooling dominates the deviation of the $\lt$ relation from its norm in
high-mass systems, we speculate that AGN heating dominates the state
of the intracluster gas in low-mass systems. Owing to the radiative
cooling of the ICM, we observe that the high-mass SCC clusters in our
sample ($\tvir > 2.5$~keV) have a higher luminosity than the high-mass
NCC clusters. The low-mass clusters ($\tvir < 2.5$~keV) with AGN are
predicted to display the reverse trend. AGN heating in these systems
becomes more dominant and causes the X-ray luminosity to decrease
relative to systems without any source of heating. This is yet to be
checked with a statistically complete sample of low-mass clusters and
groups. Our sample has only ten low-mass clusters with $\tvir <
2.5$~keV, out of which eight have strong cool-cores; low statistics of
NSCC clusters makes it difficult to test this scenario. However, this
is in line with the results obtained by \cite{Croston2005}. They
analysed a subsample of groups from the ${\it GEMS}$ group sample
\citep[Group Evolution Multiwavelength Study,][]{Osmond2004} and found
that the radio-loud groups lie on a steeper $\lt$ relation. Their
analysis, however, supports a model whereby AGN heating, instead of
the displacement of gas and the consequential decrement in the X-ray
luminosity, results in the deposition of the kinetic energy of the
radio source in the form of thermal energy, causing the gas
temperature to increase. Irrespective, we propose that it is the
combination of both of the processes, AGN heating and ICM cooling,
that causes the steepening of the $\lt$ relation.

Our results, however, seem to contradict the simulations of
\cite{Sijacki2006}. Besides self-gravity, their recipe includes
radiative cooling, photoionization, star-formation, supernovae
heating, and AGN-heating in the form of hot, radio-plasma filled
bubbles. According to their results, the effect of bubble heating is
the same for both low- and high-mass clusters and a corresponding
decrement in X-ray luminosity should be seen in CRS clusters. This is
incompatible within the framework of AGN-regulated feedback in
cool-core systems. Cool-core halos at a given temperature, because of
the nature of cooling, have an enhanced luminosity and, at the same
time as shown in \cite{Mittal2009}, undebatably favour AGN at their
centres; {\it all} SCCs have a centrally located AGN. However, this
does not imply that {\it every} SCC has an enhanced luminosity in that
low-mass SCC clusters may have a decrement because of AGN
heating. Whether there is a net increment (ICM cooling) or decrement
(AGN heating) in luminosity depends upon the scale of the system.

After applying the cool-core correction, the CC clusters show the
least intrinsic dispersion of 33\%. \cite{Pratt2009} obtained a
scatter of 26\% using luminosity and temperature in the range (0.15 to
1)$\rvir$. It is certainly plausible that the scatter decreases with
increasing radius of the region excluded for the cool-core
correction. Even though 15\%$\rvir$ is too large a region relative to
the average cooling radius of the CCs, it may be that for some of the
clusters the cooling radius does not encapsulate the total volume
affected by cooling. The slope of the CC clusters after the cool-core
correction is indeed as steep as before (note that excluding low-mass
systems with $\tvir<2.5$~keV does not make a difference). In the
context of the $\lt$ relation, this could imply that the cooling
effects are distributed over a much larger region ($\sim 0.3\rvir$)
than that suggested by the cooling radius.  There are 11 SCCs with
higher luminosities than the NCCs after the cool-core correction. A
simple calculation shows that the further decrement in the luminosity
required to match the SCC and the NCC slopes is in the range 2\% to
100\% of $\lxrc$, with a mean of around 35\%. Hence, it is plausible
that if a larger core region is subtracted for these clusters, the
slope of the SCCs would flatten. We propose that in future studies
other diagnostic regions, similar to $\rc$ and $\rcore$, should be
investigated in order to determine the optimal region over which to
make the cool-core correction.

There could be several factors contributing to the remaining scatter
in the $\lt$ relation after the cool-core correction. One such factor
is the pre-heating process in which supernovae, AGN, stars etc heat
the gas in some early epoch of cluster formation
\citep[e.g.][]{Kaiser1991,Evrard1991,Navarro1995}. The concept of
pre-heating is still not very well-understood, especially the exact
physical processes that govern it. The non-gravitational processes
associated with pre-heating are usually collectively inserted into
simulations in the form of entropy. \cite{McCarthy2004} investigated
integrated models containing radiative cooling and entropy injection
to study the global and structural properties of galaxy
clusters. Their results indicate that such models can reproduce the
observed intrinsic scatter in the scaling relations, including the
$\lt$ relation, although it may be that different levels of injection
entropy [(100 to 500)~keV~cm$^2$] are required for different mass
ranges. Lastly, self-similar clusters follow $\lx \propto f^2 \tvir^2
(1+z_{\st f})$, where $f$ is the gas mass fraction and $z_{\st f}$ is
the redshift of the cluster formation.  Hence, different epochs of
cluster formation and gas mass fraction may also contribute to the
scatter at some level.

\section{Conclusions}
\label{conc}

We have investigated the two competing physical processes that affect
the $\lt$ relation in galaxy clusters using the 64 X-ray brightest
halos in the sky. This is the largest statistically-complete sample
with high-quality X-ray and radio data. The two competing processes in
question are the cooling of the intracluster medium~(ICM) and AGN
heating. Our main results are as follows:

\begin{enumerate}
\item On cluster scales, the cooling of the intracluster medium~(ICM)
  is the dominant of the two mechanisms that shapes the $\lt$ relation
  in the strong cool-core clusters. Although they each contain a
  central radio source, the strong cool-core clusters~(SCCs) have
  enhanced luminosities.
\item SCC clusters with short cooling times~($\ct<1$~Gyr) have the
  steepest power-law index~(3.33) whereas the non-cool-core~(NCC)
  clusters with long cooling times~($\ct>7.7$~Gyr) have the shallowest
  power-law index~(2.42).
\item The intrinsic dispersion in $\lx$ ranges from a minimum of
  34.8\% for weak cool-core~(WCC) clusters~($1$~Gyr$<\ct<7.7$~Gyr) to
  maximum of 59.4\% for clusters with no central radio source~(NCRS).
\item Our results do not show that the SCCs only are responsible for
  the scatter in the $\lt$ relation. There is similar scatter for
  SCC~(51.8\%), NCC~(50.4\%), and NSCC~(47.4\%) clusters.
\item After applying the cool-core correction, the intrinsic scatter
  about the $\lt$ relation in luminosity decreases from 45.4\% to
  39.1\%. The SCCs+WCCs display the least scatter after the
  correction~(33.2\%) followed by the SCCs~(34.2\%) and then
  WCCs~(36.3\%). The NCC subset, for which the scatter is
  unchanged~(50.4\%), is the dominant source of scatter in the $\lt$
  relation after the cool-core correction.
\item The variation in the cool-core luminosity, $\lx(<\rc)$, where
  $\rc$ is the cooling radius at which $\ct=7.7$~Gyr, contributes 27\%
  to the total intrinsic scatter about the $\lt$ relation. The average
  cooling radius for SCCs+WCCs is $(0.07\pm0.03)\rvir$.
\item Owing to the nature of the sample (flux-limited), the actual
  fractions of SCC, WCC and NCC clusters may be similar. The observed
  fraction of SCCs is likely higher because of their relatively
  enhanced X-ray luminosities. The normalizations of the $\lt$
  relations of SCC, WCC, and NCC clusters were individually corrected
  for Malmquist bias, caused by the scatter, using Monte Carlo
  simulations.
\end{enumerate}

Our results suggest that on cluster scales, intracluster medium
cooling (where relevant) is more predominant than AGN heating in the
context of the $\lt$ scaling relation. Furthermore, we speculate that
AGN heating becomes increasingly important as the size of the halos
decreases and may be the dominant of the two processes on galaxy group
scales.

\begin{acknowledgements} {We thank the referee for the valuable
    comments on the manuscript and insightful
    suggestions. R.~M. acknowledges support from the Deutsche
    Forschungsgemeinschaft~(DFG) through the Schwerpunkt Program
    1177~(RE 1462/4) and T.~H.~R acknowledges support from the DFG
    through the Emmy Noether and Heisenberg research grants RE 1462/2
    and RE 1462/5.}


\end{acknowledgements}

\bibliographystyle{aa}
\bibliography{ref}

\newpage

  \begin{longtable}{|l|l|l|l|l|l|l|l|}
    \caption{Observations: (1)~Cluster name, (2) Virial temperature,
      (3) Bolometric X-ray luminosity (0.01-40)~keV, (4) Bolometric
      X-ray luminosity (0.01-40)~keV within $\rc$, the radius at which
      the gas cooling time is 7.7~Gyr, (5) $\rc$ in kpc, (6) $\rc$ in
      arcminute, (7) $\rc$ as a fraction of the virial radius, $\rvir$,
      and (8) hydrogen column densities. The columns with dashes
      ``$-$'' indicate non-cool-core clusters with $\rc=0$. }
    \label{obs}\\
    \hline
    Cluster &    k$\tvir$         &  $\lx$                                    &  $\lxrc$                             & $\rc$  & $\rc$  & $\rc/\rvir$ & $N_{\st H}$\\
                &    (keV)                & ($10^{44}$~erg~s$^{-1}$)          &  ($10^{44}$~erg~s$^{-1}$)    & (kpc)     & (arcminute)  &  &  $10^{22}$~cm$^{-2}$\\
    \hline
    \hline
    \endfirsthead
    \caption{continued.}\\
    \hline
    Cluster &    k$\tvir$         &  $\lx$                                    &  $\lxrc$                           & $\rc$  & $\rc$  & $\rc/\rvir$ &  $N_{\st H}$\\
                &    (keV)                & ($10^{44}$~erg~s$^{-1}$)          &  ($10^{44}$~erg~s$^{-1}$)    & (kpc)     & (arcminute)  &    &  $10^{22}$~cm$^{-2}$ \\
    \hline
    \endhead
    \hline
    \endfoot
\rul 
        A0085    &    6.00$^{0.11}_{-0.11}$   &      12.600$\pm$0.076   &     3.518$_{-0.020}^{0.020}$   &    123.61   &    1.95   &    0.080       &	   0.02705      \\   
\rul                                                                          											           
        A0119    &    5.73$^{0.34}_{-0.30}$   &      3.920$\pm$0.035   &      	$-$	       &     $-$     &	   $-$	 &     $-$        &	   0.03280       \\  
\rul                                                                          											           
        A0133    &    3.96$^{0.08}_{-0.10}$   &      2.810$\pm$0.023   &      1.105$_{-0.009}^{0.012}$   &    105.56   &    1.62   &    0.084       &	   0.01580	 \\  
\rul                                                                          											           
      NGC0507    &    1.44$^{0.08}_{-0.10}$   &      0.165$\pm$0.002   &      0.055$_{-0.001}^{0.001}$   &     66.75   &    3.37   &    0.088       &	   0.05560	 \\  
\rul                                                                          											           
        A0262    &    2.44$^{0.03}_{-0.04}$   &      0.793$\pm$0.030   &      0.159$_{-0.001}^{0.001}$   &     68.43   &    3.48   &    0.069       &	   0.06380	 \\  
\rul                                                                          											           
        A0400    &    2.26$^{0.10}_{-0.12}$   &      0.545$\pm$0.006   &      	$-$	       &     $-$     &	   $-$	 &     $-$        &	   0.10929	 \\  
\rul                                                                          											           
        A0399    &    6.70$^{0.14}_{-0.14}$   &      9.590$\pm$0.518   &      	$-$	       &     $-$     &	   $-$	 &     $-$        &	   0.10700	 \\  
\rul                                                                          											           
        A0401    &    8.51$^{0.34}_{-0.22}$   &      17.700$\pm$0.194   &     	$-$	       &     $-$     &	   $-$	 &     $-$        &	   0.10282	 \\  
\rul                                                                          											           
        A3112    &    4.73$^{0.12}_{-0.12}$   &      8.710$\pm$0.096   &      4.137$_{-0.032}^{0.032}$   &    127.89   &    1.51   &    0.093       &	   0.01270	 \\  
\rul                                                                          											           
      NGC1399    &    1.34$^{0.01}_{-0.01}$   &      0.053$\pm$0.003   &      0.005$_{-0.000}^{0.000}$   &     19.34   &    3.43   &    0.026       &	   0.01523	 \\  
\rul                                                                          											           
       2A0335    &    3.53$^{0.10}_{-0.13}$   &      4.080$\pm$0.033   &      3.204$_{-0.025}^{0.025}$   &    141.40   &    3.44   &    0.119       &	   0.24798	 \\  
\rul                                                                          											           
      IIIZw54    &    2.50$^{0.05}_{-0.06}$   &      0.545$\pm$0.042   &      0.065$_{-0.003}^{0.003}$   &     44.79   &    1.30   &    0.045       &	   0.14700	 \\  
\rul                                                                          											           
        A3158    &    4.99$^{0.07}_{-0.07}$   &      6.910$\pm$0.104   &      	$-$	       &     $-$     &	   $-$	 &     $-$        &	   0.01210	 \\  
\rul                                                                          											           
        A0478    &    7.34$^{0.18}_{-0.19}$   &      25.600$\pm$0.154   &     17.23$_{-0.070}^{0.054}$   &    178.08   &    1.82   &    0.104       &	   0.29276	 \\   
\rul                                                                          											                     
      NGC1550    &    1.34$^{0.01}_{-0.01}$   &      0.205$\pm$0.011   &      0.097$_{-0.003}^{0.003}$   &     72.15   &    4.84   &    0.098       &	   0.13778	 \\  
\rul                                                                          											           
      EXO0422    &    2.93$^{0.13}_{-0.12}$   &      1.760$\pm$0.109   &      0.713$_{-0.016}^{0.016}$   &     85.24   &    1.83   &    0.079       &	   0.08080	 \\  
\rul                                                                          											           
        A3266    &    9.45$^{0.35}_{-0.36}$   &      12.300$\pm$0.086   &     0.009$_{-0.001}^{0.001}$   &      9.72   &    0.14   &    0.005       &	   0.01840	 \\  
\rul                                                                          											           
        A0496    &    4.86$^{0.06}_{-0.06}$   &      3.780$\pm$0.026   &      1.458$_{-0.005}^{0.005}$   &    102.05   &    2.62   &    0.073       &	   0.04279	 \\  
\rul                                                                          											           
        A3376    &    3.80$^{0.11}_{-0.10}$   &      2.130$\pm$0.030   &      	$-$	       &     $-$     &	   $-$	 &     $-$        &	   0.04420	 \\  
\rul                                                                          											                  					    	   
        A3391    &    5.77$^{0.31}_{-0.36}$   &      2.870$\pm$0.054   &      	$-$	       &     $-$     &	   $-$	 &     $-$        &	   0.05620	 \\  
\rul                                                                          											           
       A3395s    &    4.82$^{0.26}_{-0.26}$   &      2.460$\pm$0.093   &      	$-$	       &     $-$     &	   $-$	 &     $-$        &	   0.07340	 \\  
\rul                                                                          											           
        A0576    &    4.09$^{0.09}_{-0.10}$   &      1.900$\pm$0.129   &      0.122$_{-0.003}^{0.003}$   &     45.16   &    0.99   &    0.035       &	   0.05460	 \\  
\rul                                                                          											           
        A0754    &    11.13$^{0.39}_{-0.43}$   &      6.620$\pm$0.106   &     	$-$	       &     $-$     &	   $-$	 &     $-$        &	   0.05130	 \\  
\rul                                                                          											           
        A0780    &    3.45$^{0.08}_{-0.09}$   &      6.060$\pm$0.036   &      2.889$_{-0.007}^{0.007}$   &    116.26   &    1.87   &    0.099       &	   0.05005	 \\  
\rul                                                                          											           
        A1060    &    3.16$^{0.04}_{-0.04}$   &      0.581$\pm$0.019   &      0.088$_{-0.001}^{0.001}$   &     43.41   &    2.84   &    0.039       &	   0.05030	 \\  
\rul                                                                          											           
        A1367    &    3.58$^{0.06}_{-0.06}$   &      1.130$\pm$0.009   &      	$-$	       &     $-$     &	   $-$	 &     $-$        &	   0.01719	 \\  
\rul                                                                          											           
         MKW4    &    2.01$^{0.04}_{-0.04}$   &      0.275$\pm$0.005   &      0.076$_{-0.001}^{0.001}$   &     58.50   &    2.44   &    0.065       &	   0.01710	 \\  
\rul                                                                          											           
     ZwCl1215    &    6.27$^{0.32}_{-0.29}$   &      6.250$\pm$0.081   &      	$-$	       &     $-$     &	   $-$	 &     $-$        &	   0.01760	 \\  
\rul                                                                          											           
      NGC4636    &    0.90$^{0.02}_{-0.02}$   &      0.014$\pm$0.001   &      0.006$_{-0.000}^{0.000}$   &     39.40   &    8.68   &    0.065       &	   0.01850	 \\  
\rul                                                                          											                  
        A3526    &    3.92$^{0.02}_{-0.02}$   &      1.370$\pm$0.030   &      0.366$_{-0.004}^{0.004}$   &     77.53   &    5.60   &    0.062       &	   0.08540	 \\  
\rul                                                                          											           
        A1644    &    5.09$^{0.09}_{-0.09}$   &      4.090$\pm$0.209   &      0.292$_{-0.004}^{0.005}$   &     69.69   &    1.27   &    0.049       &	   0.03990	 \\  
\rul                                                                          											           
        A1650    &    5.81$^{0.06}_{-0.07}$   &      9.550$\pm$0.630   &      2.333$_{-0.012}^{0.012}$   &     94.52   &    1.01   &    0.062       &	   0.01300	 \\  
\rul                                                                          											           
        A1651    &    6.34$^{0.27}_{-0.27}$   &      9.860$\pm$0.118   &      2.475$_{-0.052}^{0.050}$   &     97.56   &    1.03   &    0.061       &	   0.01460	 \\  
\rul                                                                          											           
        A1656    &    9.15$^{0.17}_{-0.17}$   &      11.100$\pm$0.156   &     	$-$	       &     $-$     &	   $-$	 &     $-$        &	   0.00767	 \\  
\rul                                                                          											           
      NGC5044    &    1.22$^{0.03}_{-0.04}$   &      0.115$\pm$0.001   &      0.089$_{-0.002}^{0.002}$   &     85.46   &    8.06   &    0.122       &	   0.05060	 \\  
\rul                                                                          											           
        A1736    &    3.12$^{0.11}_{-0.12}$   &      2.920$\pm$0.184   &      	$-$	       &     $-$     &	   $-$	 &     $-$        &	   0.04315	 \\  
\rul                                                                          											           
        A3558    &    4.95$^{0.13}_{-0.15}$   &      7.620$\pm$0.038   &      0.670$_{-0.012}^{0.012}$   &     67.96   &    1.22   &    0.048       &	   0.03997	 \\  
\rul                                                                          											           
        A3562    &    4.43$^{0.21}_{-0.16}$   &      3.350$\pm$0.030   &      0.150$_{-0.006}^{0.006}$   &     46.69   &    0.82   &    0.035       &	   0.03999	 \\  
\rul                                                                          											                  					    	   
        A3571    &    7.00$^{0.13}_{-0.12}$   &      10.200$\pm$0.071   &     2.556$_{-0.017}^{0.017}$   &     65.49   &    1.43   &    0.039       &	   0.04382	 \\  
\rul                                                                          											           
        A1795    &    6.08$^{0.07}_{-0.07}$   &      14.800$\pm$0.044   &     6.095$_{-0.018}^{0.016}$   &    137.63   &    1.93   &    0.088       &	   0.01015	 \\  
\rul                                                                          											           
        A3581    &    1.97$^{0.07}_{-0.07}$   &      0.544$\pm$0.017   &      0.281$_{-0.005}^{0.005}$   &     87.82   &    3.19   &    0.099       &	   0.04310	 \\  
\rul                                                                          											           
         MKW8    &    3.00$^{0.12}_{-0.12}$   &      0.692$\pm$0.058   &      	$-$	       &     $-$     &	   $-$	 &     $-$        &	   0.02340	 \\  
\rul                                                                          											           
      RXJ1504    &    9.53$^{1.39}_{-1.16}$   &      112.000$\pm$1.120   &    53.875$_{-0.249}^{0.249}$  &    234.84   &    1.13   &    0.120       &	   0.06080	 \\   
\rul                                                                          											           
        A2029    &    8.26$^{0.09}_{-0.09}$   &      27.600$\pm$0.166   &     12.076$_{-0.026}^{0.026}$  &    142.19   &    1.64   &    0.078       &	   0.03214	 \\   
\rul                                                                          											                 
        A2052    &    3.35$^{0.02}_{-0.02}$   &      2.180$\pm$0.022   &      1.052$_{-0.005}^{0.005}$   &    104.70   &    2.50   &    0.090       &	   0.02680	 \\  
\rul                                                                          											           
        MKW3S    &    3.90$^{0.09}_{-0.09}$   &      2.690$\pm$0.027   &      1.214$_{-0.009}^{0.009}$   &     92.47   &    1.76   &    0.074       &	   0.02860	 \\  
\rul                                                                          											           
        A2065    &    5.40$^{0.20}_{-0.11}$   &      6.650$\pm$0.406   &      1.061$_{-0.012}^{0.013}$   &     85.95   &    1.05   &    0.058       &	   0.03285	 \\  
\rul                                                                          											           
        A2063    &    3.77$^{0.06}_{-0.06}$   &      2.060$\pm$0.027   &      0.349$_{-0.004}^{0.004}$   &     64.93   &    1.58   &    0.053       &	   0.02730	 \\  
\rul                                                                          											           
        A2142    &    8.40$^{1.01}_{-0.76}$   &      36.100$\pm$0.324   &     7.804$_{-0.044}^{0.043}$   &    125.73   &    1.25   &    0.068       &	   0.04151	 \\  
\rul                                                                          											           
        A2147    &    4.07$^{0.11}_{-0.12}$   &      3.100$\pm$0.099   &      	$-$	       &     $-$     &	   $-$	 &     $-$        &	   0.02823	 \\  
\rul                                                                          											           
        A2163    &    15.91$^{0.81}_{-0.81}$   &      75.500$\pm$1.130   &    	$-$	       &     $-$     &	   $-$	 &     $-$        &	   0.11765	 \\  
\rul                                                                          											           
        A2199    &    4.37$^{0.07}_{-0.07}$   &      4.030$\pm$0.072   &      1.644$_{-0.008}^{0.008}$   &     97.99   &    2.74   &    0.074       &	   0.00857	 \\  
\rul                                                                          											           
        A2204    &    8.92$^{0.72}_{-0.61}$   &      39.600$\pm$0.634   &     23.886$_{-0.237}^{0.177}$  &    171.08   &    1.09   &    0.090       &	   0.05921	 \\   
\rul                                                                          											           
        A2244    &    5.78$^{0.10}_{-0.11}$   &      11.700$\pm$0.246   &     3.758$_{-0.022}^{0.025}$   &    119.86   &    1.13   &    0.079       &	   0.01908	 \\  
\rul                                                                          											                 
        A2256    &    7.61$^{0.65}_{-0.63}$   &      11.200$\pm$0.156   &     	$-$	       &     $-$     &	   $-$	 &     $-$        &	   0.04566	 \\  
\rul                                                                          											           
        A2255    &    5.81$^{0.19}_{-0.20}$   &      7.510$\pm$0.090   &      	$-$	       &     $-$     &	   $-$	 &     $-$        &	   0.02493	 \\  
\rul                                                                          											           
        A3667    &    6.39$^{0.04}_{-0.04}$   &      12.600$\pm$0.088   &     0.015$_{-0.000}^{0.000}$   &     12.39   &    0.19   &    0.008       &	   0.04590	 \\  
\rul                                                                          											                  					    	   
        S1101    &    2.57$^{0.12}_{-0.13}$   &      3.130$\pm$0.028   &      2.368$_{-0.026}^{0.027}$   &    171.11   &    2.57   &    0.169       &	   0.01050	 \\  
\rul                                                                          											           
        A2589    &    3.89$^{0.05}_{-0.05}$   &      1.780$\pm$0.023   &      0.371$_{-0.003}^{0.003}$   &     72.72   &    1.50   &    0.058       &	   0.02880	 \\  
\rul                                                                          											           
        A2597    &    4.05$^{0.07}_{-0.07}$   &      7.320$\pm$0.088   &      4.429$_{-0.016}^{0.016}$   &    125.34   &    1.32   &    0.098       &	   0.02200	 \\  
\rul                                                                          											           
        A2634    &    3.19$^{0.11}_{-0.11}$   &      0.946$\pm$0.015   &      0.004$_{-0.000}^{0.000}$   &     12.03   &    0.32   &    0.011       &	   0.05140	 \\  
\rul                                                                          											           
        A2657    &    3.52$^{0.12}_{-0.11}$   &      1.640$\pm$0.015   &      0.186$_{-0.006}^{0.006}$   &     51.03   &    1.08   &    0.043       &	   0.05915	 \\  
\rul                                                                          											           
        A4038    &    3.14$^{0.03}_{-0.04}$   &      1.900$\pm$0.025   &      0.614$_{-0.013}^{0.013}$   &     82.24   &    2.31   &    0.073       &	   0.01540	 \\  
\rul                                                                          											           
        A4059    &    4.22$^{0.03}_{-0.03}$   &      3.140$\pm$0.041   &      1.169$_{-0.005}^{0.005}$   &    109.85   &    1.99   &    0.084       &	   0.01058       \\  
    \hline\hline
  \end{longtable}

\end{document}